\documentclass [11pt]{article}

\usepackage{amsmath,amssymb,epsfig,cite,color,verbatim}
\usepackage{graphics,float}
\usepackage{overpic}
\usepackage{physics}
\usepackage{mathrsfs}
\usepackage{slashed,bbold}
 \usepackage{multicol, array}
\usepackage{cancel, soul}
\usepackage{soul}
\usepackage{hyperref}
\usepackage{multirow}
\usepackage{xcolor}

\usepackage[normalem]{ulem}
\usepackage{amsmath}
\newcommand{\stkout}[1]{\ifmmode\text{\sout{\ensuremath{#1}}}\else\sout{#1}\fi}

\numberwithin{equation}{section}

\title{ Quantum phases at high chemical potential in 2-flavor matrix-QC$_2$D}
\date{\today}
\author{Nirmalendu Acharyya$^1$\footnote{nirmalendu@iitbbs.ac.in}, Prasanjit Aich$^{2}$\footnote{prasanjita@iisc.ac.in}, Arkajyoti Bandyopadhyay$^1$\footnote{s22ph09003@iitbbs.ac.in} and Sachindeo~Vaidya$^2$\footnote{vaidya@iisc.ac.in} \\
${}^1${\small School of Basic Sciences, Indian Institute of Technology Bhubaneswar, Jatni, Khurda, Odisha 752050, India}\\
${}^2${\small Centre for High Energy Physics,  Indian Institute of Science, Bengaluru, 560012, India}\\
}

\begin{document}

\maketitle

\begin{abstract} 
We investigate the matrix model of  two-color two-flavor QCD (matrix-QCD$_{2,2}$) in regimes with large baryon ($\mu_{_B}$),  isospin ($\mu_{_I}$), and/or chiral ($c$) chemical potentials. In these regimes,  the Hamiltonian simplifies  considerably, making it possible to investigate the ground state for intermediate and strong Yang-Mills coupling. By diagonalizing the Hamiltonian using the variational techniques, we show that in regimes where $\mu_{_B}$  and $c$ (or $\mu_{_B}$ and $\mu_{_I}$) dominate, tuning the remaining parameters leads to quantum phase transitions (QPTs). These transitions form a complex web of phases, each of which has a ground state uniquely labelled by baryon number $B$ and isospin $I$. 
 Several of these phases are LOFF-like, characterized by a ground state carrying non-zero spin and hence spontaneously breaking rotational symmetry.  These results are consistent with older effective field theory predictions by Splittorff-Son-Stephanov \cite{Splittorff:2000mm}.  The fermionic content of these LOFF-like ground states consists of  spin-1 di-(anti-) quarks  which are analogous to Cooper pairs. We  compute the spin-fraction carried by the quarks and find that it constitutes a significant portion -- in some cases nearly the entirety -- of the total spin.

\end{abstract}

\section{Introduction}

Understanding strongly interacting matter at nonzero baryon and isospin densities is a central problem in particle physics. Studying quantum chromodynamics (QCD) in these regimes is relevant to diverse physical settings, like the interior of neutron stars, supernovae, early universe, and strongly interacting matter created in heavy-ion collisions \cite{Weber:2004kj, Cheng:2009zr,Abdikamalov:2008df, Sagert:2008ka, Alford:2007xm, Gupta:2011wh, Stephanov:1999zu, Mohanty:2011ki}. 

 In presence of non-zero baryon and isospin chemical potentials ($\mu_{_B}$ and $\mu_{_I}$ respectively), several low-energy effective models predict a rich phase structure. Among these, there are  several exotic phases   \cite{He:2006tn, Sun:2007fc, Mu:2010zz} and in  particular,  inhomogeneous superconducting phases \cite{Son:2000xc}.

However, direct numerical verifications of these theoretical predictions using lattice QCD simulations with finite baryon density  remain  extremely challenging. 
The biggest obstacle for lattice QCD originates from the complex fermion determinant when $\mu_{_B} \neq 0$ \cite{Engels:1985ts, Karsch:1985cb}, leading to the infamous  sign problem \cite{Muroya:2003qs, Splittorff:2007ck, deForcrand:2009zkb}.  In this context, QC$_2$D -- two-color QCD has gained substantial attention because  it avoids the technical hurdle of the sign problem: as  the irreps  of $SU(2)$ are (pseudo-)real, the fermion determinant remains real even when $\mu_{_B}\neq0$ and lattice simulations of such situations are feasible \cite{Hands:1999md,Kogut:1999ke,Kogut:2001na}. 

But QC$_2$D has some notable differences compared to three-color QCD. For instance, here the baryons (in fact, all hadrons) have integer spin. 
Nevertheless, QC$_2$D still captures  several key features of the three-color QCD and hence provides useful insights  \cite{Kogut:2000ek,Nishida:2003uj,Wirstam:2002be,Splittorff:2002xn}.   With non-zero baryon and isospin chemical potentials, it exhibits  a rich phase structure,  including diquark condensed phase, superfluid phases, and crossovers between different dense-matter regimes \cite{Kogut:1999ke,Kogut:2002cm,Hands:2010gd,Cotter:2012mb, Kojo:2021hqh, Astrakhantsev:2020tdl,Begun:2022bxj,Iida:2024irv}.  Thus QC$_2$D is interesting in its own right and has been investigated using effective field theories  \cite{Kogut:1999ke,Kogut:2000ek,Splittorff:2002xn, Splittorff:2000mm,Andersen:2010vu, Suenaga:2025} and lattice simulations \cite{Nakamura:1984, Hands:1999md, Kogut:2001na, Kogut:2002cm, Hands:2010gd, Cotter:2012mb, Braguta:2015cta, Astrakhantsev:2020tdl, Begun:2022bxj, Braguta:2023yhd, Iida:2024irv, Muroya:2002ry, Kogut:2003ju, Hasan:2026ggs, Kogut:2001if}. 
In particular, in 2-flavor QC$_2$D with  non-zero $\mu_{_B}$ and $\mu_{_I}$, there are predictions of inhomogeneous phases, where a di-(anti-)quark condensate breaks the translational and rotational symmetries \cite{Splittorff:2000mm, Fukushima:2007bj}.  These phases are similar to the LOFF phases \cite{Larkin:1964wok, Fulde:1964zz}. 
 
The matrix model, first proposed for pure gauge theory in \cite{Balachandran:2014iya,Balachandran:2014voa} and later extended to include quarks, provides a simplified platform for both analytic studies \cite{Pandey:2016hat, Acharyya:2021egi} and quantitative predictions in the low-energy regime of QCD \cite{ Acharyya:2017uhl, Acharyya:2024pqj, Acharyya:2026uhx, Acharyya:2016fcn, Pandey:2019dbp}. Here we investigate the matrix model of  two-flavor QC$_2$D, or matrix-QCD$_{2,2}$, in  the presence of baryon, isospin and chiral ($c$) chemical potentials. Specifically, we explore the regimes  where $\mu_{_B}$, $c$ and/or $\mu_{_I}$ are large. We show that in these regions of the parameter space, the ground state of the theory is characterized by the (non-trivial) baryon number $B$ and isospin $I$.  On varying these parameters, quantum phase transitions (QPTs) happen because of level crossings and across such QPTs, $B$ and $I$ change discontinuously.  As functions of $\Delta \equiv c-\mu_{_B}$ and $\mu_{_I}$, these phases form a complex web.    In particular, we show that for specific values of $\Delta$ and $\mu_{_I}$,  the ground state has non-zero spin and hence spontaneously breaks rotational symmetry.  Additionally, the ground state has substantial quark content in the form of di-(anti-)quarks with non-zero isospin.

This is an appropriate juncture to recall features of QC$_2$D with non-zero $\mu_{_B}$ and/or $\mu_{_I}$.  In the presence of such chemical potentials, it is possible to form di-quarks (similar to Cooper pairs) with non-zero center-of-mass momentum, which can condense to form a new phase \cite{Splittorff:2000mm, Fukushima:2007bj, Brauner:2019rjg, Braguta:2023yhd}.  
This phase spontaneously breaks the translational and rotational invariance.  Our analysis here is directly comparable to \cite{Splittorff:2000mm}, where it has been argued that  when $\mu_{_B}$ and/or $\mu_{_I}$ are  very large, the di-quark condensate  has non-zero isospin and carries orbital angular momentum or spin. In contrast, the complementary situation of small $\mu_{_B}$ and $\mu_{_I}$ is discussed in \cite{Fukushima:2007bj}, where the  condensate has $I=0$ and does not carry any angular momentum.  While the arguments for the existence of these phases are well-motivated,  their direct observations in  numerical simulations remain elusive.

Of course the matrix model cannot capture the inhomogeneous nature of the LOFF phase, as it has no degrees of freedom corresponding to spatial variation.  Nonetheless, our analysis shows  that in certain regimes of chemical potentials, the ground state has non-zero isospin, and is dominated by di-(anti-)quarks carrying non-zero spin.   It is in this sense that these phases of the matrix model mimic the LOFF phases in QC$_2$D  \cite{Splittorff:2000mm}. We have investigated these LOFF windows for both intermediate and strong  coupling.
 To quantify the significance of the di-quark content, we analyze the spin-fractions of the ground state carried by the glue and the quarks and show that the total spin of the ground state  is almost always dominated by the  quarks. 
In section 2, we present the Hamiltonian of the  model, discuss its symmetries, and the limits of $\mu_B\to \infty$ and/or  $\mu_I\to \infty$. We derive the effective Hamiltonian of the theory in these regimes and  apply the variational method to study the properties of the ground state. Our main results are presented in section 3. We show that in these parameter regimes, there are level crossings, leading to QPTs.  We present the complete phase diagram in the large $\mu_{_B}$, $\mu_{_I}$ and/or $c$ limits.    Several  of these phases are analogous to the LOFF phases observed in QC$_2$D \cite{Splittorff:2000mm}. For these LOFF-like phases, we compute the spin-fraction carried by the fermionic di-quark states and establish their role as the analogues of the Cooper pairs in these phases.

\section{Hamiltonian}  
The construction of the matrix model involves in a fundamental way the pullback of the Maurer-Cartan form of $SU(2)$ (or $SU(N)$) to the spatial three-sphere $S^3$ of radius $R$ \cite{Singer:1978dk, Narasimhan:1979kf}.  
Here, the glue (gauge degrees of freedom) is a $3\times3$ real matrix $M_{ia}$, where $i$ is  the spatial index and $a$ the color index.  
The dynamical variables  of the pure gauge theory are  $M_{ia}$ and their conjugate momenta $P_{ia}\equiv -\,i\,\frac{\partial}{\partial M_{ia}}$.  The chromoelectric field  is $E_{ia}= P_{ia}$,  while the chromomagnetic field is  $
B_{ia} = -\,\frac{1}{R} M_{ia}  + \frac{1}{2}\,\epsilon_{ijk} \epsilon_{abc} M_{jb} M_{kc}$. The pure Yang-Mills Hamiltonian is given by 
\begin{eqnarray}
    H_{YM} =\left[\frac{g^2}{2 R^3} E_{ia} E_{ia} + \frac{R^3}{2g^2} B_{ia} B_{ia}\right]
\end{eqnarray}
where  $g$ is the Yang–Mills coupling. 

The quarks transforming in the fundamental representation $\psi_{\alpha A f}$ are Grassmann-valued matrices which depend only on time. In the Weyl representation, they can be represented as $\psi_{\alpha A f} = (b_{\alpha A f}, \,\,\,
d_{\alpha A f}^\dagger)^T$ where the $b_{\alpha Af}$'s and $d_{\alpha Af}$'s satisfy $\{b_{\alpha A f}, b_{\beta B f'}^\dagger\}=R^{-3}\delta_{\alpha \beta} \delta_{AB} \delta_{ff'}$ and  $\{d_{\alpha A f}, d_{\beta B f'}^\dagger\}=R^{-3} \delta_{\alpha \beta} \delta_{AB} \delta_{ff'}$. 
These quarks interact with the glue via minimal coupling \cite{Pandey:2019dbp}: $H_{int} \equiv  R^3 (b^\dagger_{\alpha A f} \sigma^i_{\alpha \beta} T^a_{AB}b_{\beta Bf} - d_{\alpha Af} \sigma^i_{\alpha \beta} T^a_{AB}d^\dagger_{\beta Bf}) M_{ia}$. In general, these quarks can be massive:  the mass term in the Hamiltonian is $ H_m \equiv  \widetilde{m}  R^3h_m$ where $h_m\equiv (b^\dagger_{\alpha A f} d^\dagger_{\alpha A f} +d_{\alpha A f} b_{\alpha A f})$ and $\widetilde{m} \geq 0$ is the quark mass.  We also will study the theory  with non-zero  baryon number and isospin   chemical potentials. The corresponding terms in the Hamiltonian are $(2\widetilde{\mu}_{_{B}}  R^3 Q_B)$ and $ (2\widetilde{\mu}_{_{I}} R^3 Q_I)$, where 
$Q_B=  \frac{1}{2}(b^\dagger_{\alpha A f} b _{\alpha A f} - d^\dagger_{\alpha A f} d _{\alpha A f})$ and $Q_I=  \frac{1}{2}(b^\dagger_{\alpha A f}\tau^3_{ff'} b _{\alpha A f'} - d^\dagger_{\alpha A f} \tau^3_{ff'} d _{\alpha A f'})$.  
The  parameters ${\mu}_{_{B}},{\mu}_{_{I}}\in\mathbb{R}$ are the baryon number and isospin chemical potentials. Further, we add the chiral chemical potential $ 2\widetilde{c}   R^3 Q_c,$ where  $Q_c\equiv \frac{1}{2}(b^\dagger_{\alpha A f} b _{\alpha A f} - d_{\alpha A f} d^\dagger _{\alpha A f})$ \cite{Acharyya:2024pqj, Braguta:2015owi,Braguta:2016aov}.

We are interested in the intermediate and strong coupling regimes: $\nu \equiv g^{-2/3} \leq 1.$
In this regime, it  is convenient to work with dimensionless parameters.  In order to do so, we rescale  $M_{ia}\rightarrow R^{-1} g^{2/3}M_{ia}$, $P_{ia}\rightarrow R\,g^{-2/3}P_{ia}$, $b_{\alpha A f} \to R^{-\frac{3}{2}}b_{\alpha A f}$ and $d_{\alpha A f} \to R^{-\frac{3}{2}}d_{\alpha A f}$. Under this rescaling, the Hamiltonian becomes
\begin{eqnarray}
H&=& e_0\Big[H_0+ m \,h_m +  2 {\mu}_{_{B}}\,  Q_B +  2{\mu}_{_{I}} \,Q_I +  2{c}\, Q_c \Big] \label{Ham_1} \\
H_0&=&   \frac{1}{2} P_{ia} P_{ia} +\frac{\nu^2}{2} M_{ia} M_{ia} - \frac{\nu}{2} \epsilon_{ijk}\epsilon_{abc} M_{ia} M_{jb} M_{kc}+ \frac{1}{4} \epsilon_{abc} \epsilon_{ade} M_{ib} M_{jc} M_{id} M_{je}+ \nonumber \\
&& \,\,\,\, \,\,  (b^\dagger_{\alpha A f} \sigma^i_{\alpha \beta} T^a_{AB}b_{\beta Bf} - d_{\alpha Af} \sigma^i_{\alpha \beta} T^a_{AB}d^\dagger_{\beta Bf}) M_{ia}\label{Ham_2}
\end{eqnarray} 
where $e_0 \equiv (\nu R)^{-1}$ is the energy scale of the theory. The parameters ${m}$, ${\mu}_{_{B}}$, ${\mu}_{_{I}}$ and ${c}$ are dimensionless: 
\begin{eqnarray}
{m}\equiv \frac{\widetilde{m}}{e_0},\quad\quad {\mu}_{_{B}}\equiv \frac{\widetilde{\mu}_{_{B}}}{e_0}, \quad\quad {\mu}_{_{I}}\equiv \frac{\widetilde{\mu}_{_{I}}}{e_0},  \quad\quad c\equiv \frac{\widetilde{c}}{e_0}. 
\end{eqnarray}
Using (\ref{Ham_1}), we probe the coupling range $1 \geq \nu \geq 0$. The $\nu\simeq 1$ is the intermediate coupling region, while  $\nu =0$ is the extreme strong coupling limit.

In the rescaled variables, the operators $Q_B$, $Q_I$, $h_m$ and $Q_c$ are all dimensionless operators. It is easy to check that the Hamiltonian commutes with $Q_B$ and $Q_I$, which  generate a $U(1)_B \times U(1)_I$ symmetry.  On the other hand,  $Q_c$ commutes with $H$ only if the quarks are massless. However, the $U(1)_A$ symmetry generated by $Q_c$ when $m=0$ survives only at the classical level. In the quantum theory with massless quarks, $U(1)_A$  is broken by axial anomaly \cite{Acharyya:2021egi}. For the massive theory, $U(1)_A$  is  explicitly broken. Thus $Q_c$ is not a generator of a symmetry, rather it is an observable. On the other hand, the eigenvalues of $Q_B$ and $Q_I$ are the baryon number  $B$ and isospin $I$, which are additional quantum numbers for the states (besides the spin quantum numbers).

\subsection{Effective Hamiltonian for large chemical potentials }
Defining the quark and antiquark number operators $N_{b_f}\equiv b_{\alpha A f}^\dagger b_{\alpha A f}$ and $ N_{d_f}\equiv d_{\alpha A f}^\dagger d_{\alpha A f}$, 
we can recast the Hamiltonian (\ref{Ham_1}) as: 
\begin{eqnarray}
H= e_0\Big[H_0 + m \, h_m + ({c}+{\mu}_{_{B}}+{\mu}_{_{I}} )N_{b_1} +({c}+{\mu}_{_{B}}-{\mu}_{_{I}} )N_{b_2} + ({c}-{\mu}_{_{B}}-{\mu}_{_{I}} )N_{d_1}+ ({c}-{\mu}_{_{B}}+{\mu}_{_{I}} )N_{d_2}\Big]
\end{eqnarray}
In the limit ${\mu}_{_{B}} \to \infty$ and ${c} \to \infty$ with finite $\Delta \equiv {c}- {\mu}_{_{B}}$, the low-energy states $|\cdot\rangle$ have $N_{b_1}|\cdot\rangle =0 = N_{b_2}|\cdot\rangle$. 
Thus, the low-energy states are purely antiquark states with $0\leq N_{d_f}\leq 4$.  The states with non-zero $N_{b_f}$ are of much higher energies.

As $h_m$ creates (or destroys) a quark-antiquark pair, its matrix element is non-zero only between states with different $N_{b_f}$. But as the energy gap between two states with different $N_{b_f}$ is very large, these matrix elements are  insignificant compared to this gap, provided that the quark mass ${m}\ll {\mu}_{_{B}}$.  As a result,  ${m}$ does not affect the spectra in this limit.

Thus in this limit, the effective Hamiltonian thus reduces to 
\begin{eqnarray}
    H \simeq  e_0\Big[H_0 + \Delta(N_{d_1}+N_{d_2})-{\mu}_{_{I}} (N_{d_1}-N_{d_2}) \Big], \label{approx_1}
\end{eqnarray}
where $\Delta$ and ${\mu}_{_{I}}$ remain finite (positive or negative). 
The baryon number and the isospin of the low-energy states are $B \in [-4,0]$ and $I\in [-2,2]$.

If instead, we consider the limit ${\mu}_{_{B}} \to -\infty$ and ${c} \to \infty$ with finite $\Delta \equiv {c}+ {\mu}_{_{B}}$, the low-energy states $|\cdot\rangle$ have $N_{d_1}|\cdot\rangle =0 = N_{d_2}|\cdot\rangle$. 
The low-energy states are purely quark states with $0\leq N_{b_f}\leq 4$ and the  Hamiltonian in this case is 
\begin{eqnarray}
    H \simeq  e_0\Big[H_0 + \Delta(N_{b_1}+N_{b_2})+{\mu}_{_{I}} (N_{b_1}-N_{b_2}) \Big].  \label{approx_2}
\end{eqnarray}
The baryon number and the isospin of the low-energy states are $B \in [0,4]$ and $I\in [-2,2]$.

Similarly, we can also consider the limits  ${\mu}_{_{B}} \to \infty, c \to -\infty$ and ${\mu}_{_{B}} \to -\infty, c \to -\infty$. The effective Hamiltonian for these cases can be obtained similar to above. However, since this corresponds to replacing quarks by anti-quarks and/or particles by holes, the results for these cases can be easily read off from the previous cases.

A different interesting limit is  ${\mu}_{_{B}} \to \infty$  and ${\mu}_{_{I}} \to \infty$ with finite ${\mu}_{_{BI}} \equiv {\mu}_{_{B}}-{\mu}_{_{I}}$ and $c$. Here, the low-energy states have $N_{b_1}=0$ and $N_{d_1}=4$. As these states can have both quarks and antiquarks ($0\leq N_{b_2}\leq 4$ and $0\leq N_{d_2}\leq 4$), the quark mass term remains important in this limit.

Thus the  Hamiltonian (after removing the constant $-4 ({\mu}_{_{B}}+{\mu}_{_{I}}-c)$) in this limit is 
\begin{eqnarray}
    H \simeq  e_0\Big[H_0 + {m} \, h_m  + {c} (N_{b_2} + N_{d_2}) + {\mu}_{_{BI}} (N_{b_2} - N_{d_2}) \Big]. 
\end{eqnarray}
The baryon number and the isospin of these states are related: 
\begin{eqnarray}
B=-2 + \ell, \quad\quad I=-2-\ell, \quad \quad \text{where } -2\leq \ell \leq 2.  \label{N_B_I_reln}
\end{eqnarray}

\begin{figure}[ht!]
\begin{center} 
\includegraphics[width=17.75cm]{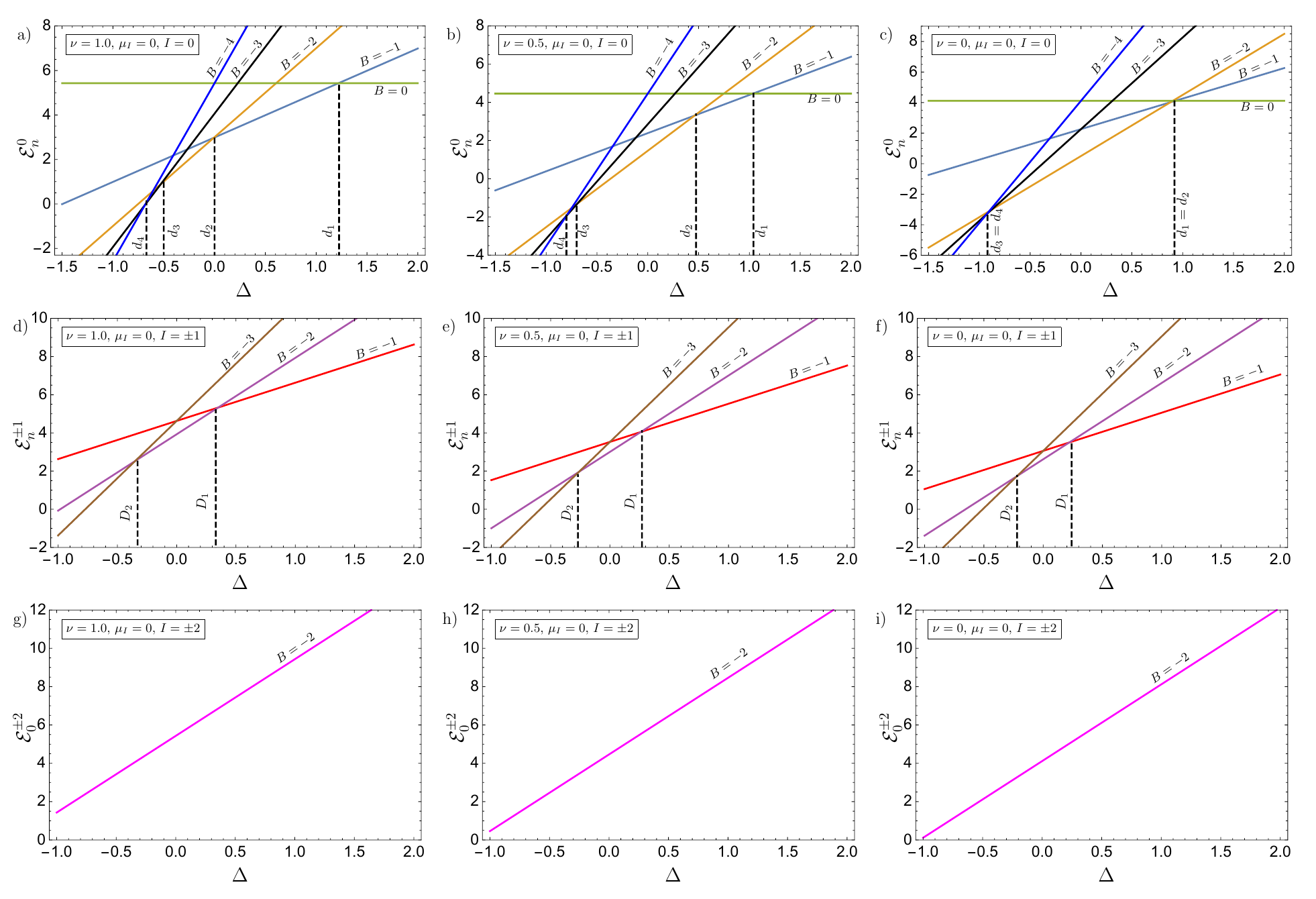}
\caption{Low-lying energy eigenvalues $\mathcal{E}_n^I$ as a function of $\Delta$ when ${\mu}_{_{I}} = 0$
for different values of the coupling $\nu$. a-c) $I = 0$. d-f) $I=\pm 1$, g-i) $I=\pm2$. 
The colored curves denote states with different baryon number $B = 0,-1,-2,-3,-4$.
The dashed vertical lines mark the values of $\Delta$ where level crossings
occur among the lightest states.} \label{Fig2_sebsec_1}
\end{center}
\end{figure}

\begin{figure}[ht!]
\begin{center} 
\includegraphics[width=15cm]{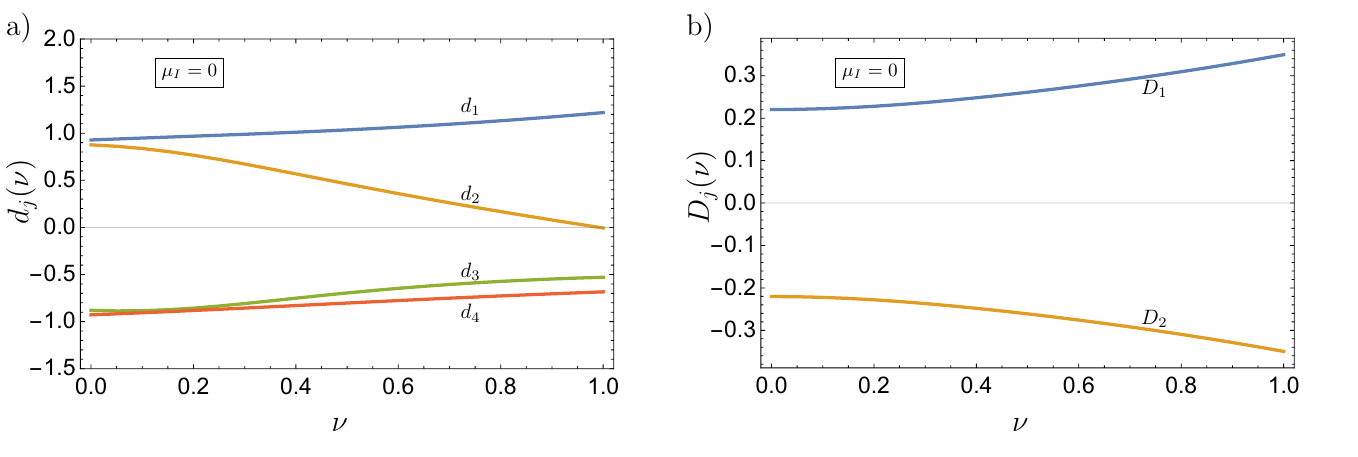}
\caption{The locations of the level crossings in the lightest states of different $I$-sectors when ${\mu}_{_{I}}=0$ as a function of $\nu$. a) $d_j(\nu)$ with $j=1,2,3,4$ -- the locations  of level crossings in the lightest $I=0$ state. b) $D_j(\nu)$ with $j=1,2$ -- the locations  of level crossings in the lightest $I=\pm 1$ state.}\label{Fig_di_Di}
\end{center}
\end{figure}
\section{Results}
These color-singlet energy eigenstates can be labeled by their spin $J$, baryon number $B$, and isospin $I$. 
We are interested in the low-lying colorless eigenstates (see Appendix~\ref{app_sym}) of the Hamiltonian (\ref{Ham_1}) when the baryon (and isospin) and/or chiral chemical potentials are large. We explore both the intermediate ($\nu \simeq 1$) and strong ($\nu = 0$) coupling regimes.

For the Hamiltonian (\ref{Ham_1}),  exact diagonalization using analytic methods is not possible. Here, we will use the variational technique to construct the colorless eigenstates and estimate their energies. The numerical strategy for the variational computation is well known \cite{Acharyya:2024pqj, Acharyya:2026uhx}.

For the purpose of the numerical simulations, we expand the bosonic wavefunctions in the basis of harmonic oscillator energy eigenstates. We truncate these expansions with an upper limit on the number of  harmonic oscillator quanta $N_b$.  
We progressively increase $N_b$ till we reach convergence in the energy eigenvalues.  For most of our work, we find that the numerical energies converge well with $N_b \sim 16$ (the data for $N_b=16$ and $N_b=18$ are practically indistinguishable).  For certain special observables, the data  do not converge: as we discuss below, this is not due  to  the failure of the numerical recipe but rather due  to the emergence of an interesting physical situation. 

\begin{figure}[ht!]
\begin{center} 
\includegraphics[width=18cm]{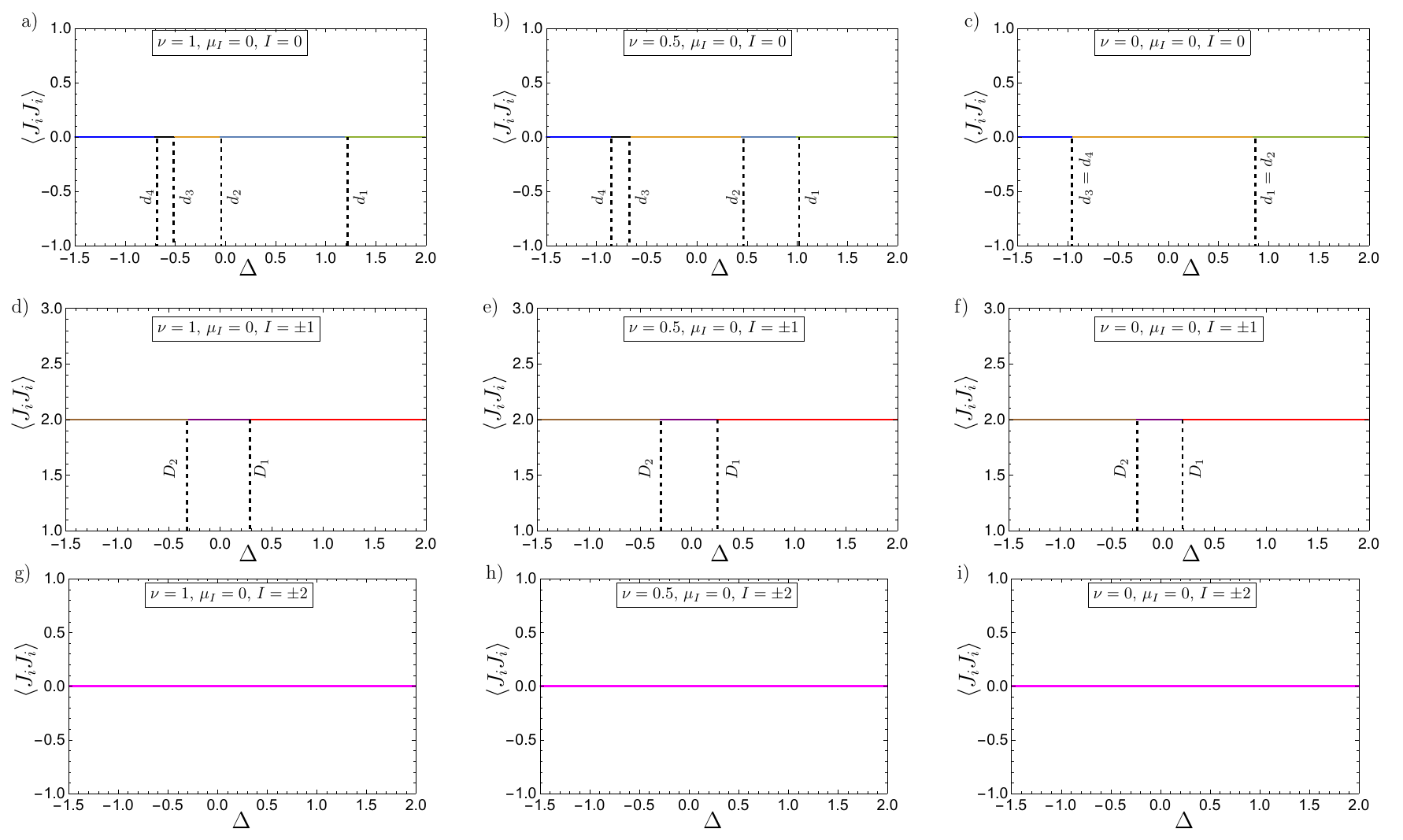}
\caption{ $\langle J_i J_i\rangle $  in the lightest states with different $I$ as a function of $\Delta$ at ${\mu}_{_{I}}=0$ for various values of $\nu$. The dashed  lines indicate the locations of the level crossings.} \label{Fig7_sebsec_1}
\end{center}
\end{figure}

\subsection{$|{\mu}_{_{B}}|\to \infty$ and $|{c}| \to \infty$ }\label{sec4_1}
Let us first consider the regime in which ${\mu}_{_{B}}\to\infty$ and ${c}\to\infty$, while ${\mu}_{_{I}}$ and the difference $\Delta \equiv {c}-{\mu}_{_{B}}$ remain finite.  
The other signs of ${\mu}_{_{B}}$ and/or ${c}$  lead to obvious changes in the baryon number and isospin of the ground state (see (\ref{approx_1})-(\ref{approx_2})), and it is not necessary to study these cases separately.

In this regime, the mass ${m}$ is an irrelevant parameter, as we argued earlier.   
As we are primarily interested in the ground state, it is sufficient to only consider the low-lying spin-0 and spin-1  
states with baryon number $B\in[-4,0]$ and isospin $I\in[-2,2]$. The states in the other sectors (with different spin, baryon number and isospin) are not relevant for this discussion as their energies are significantly higher.

\begin{figure}[ht!]
\begin{center} 
\includegraphics[width=18cm]{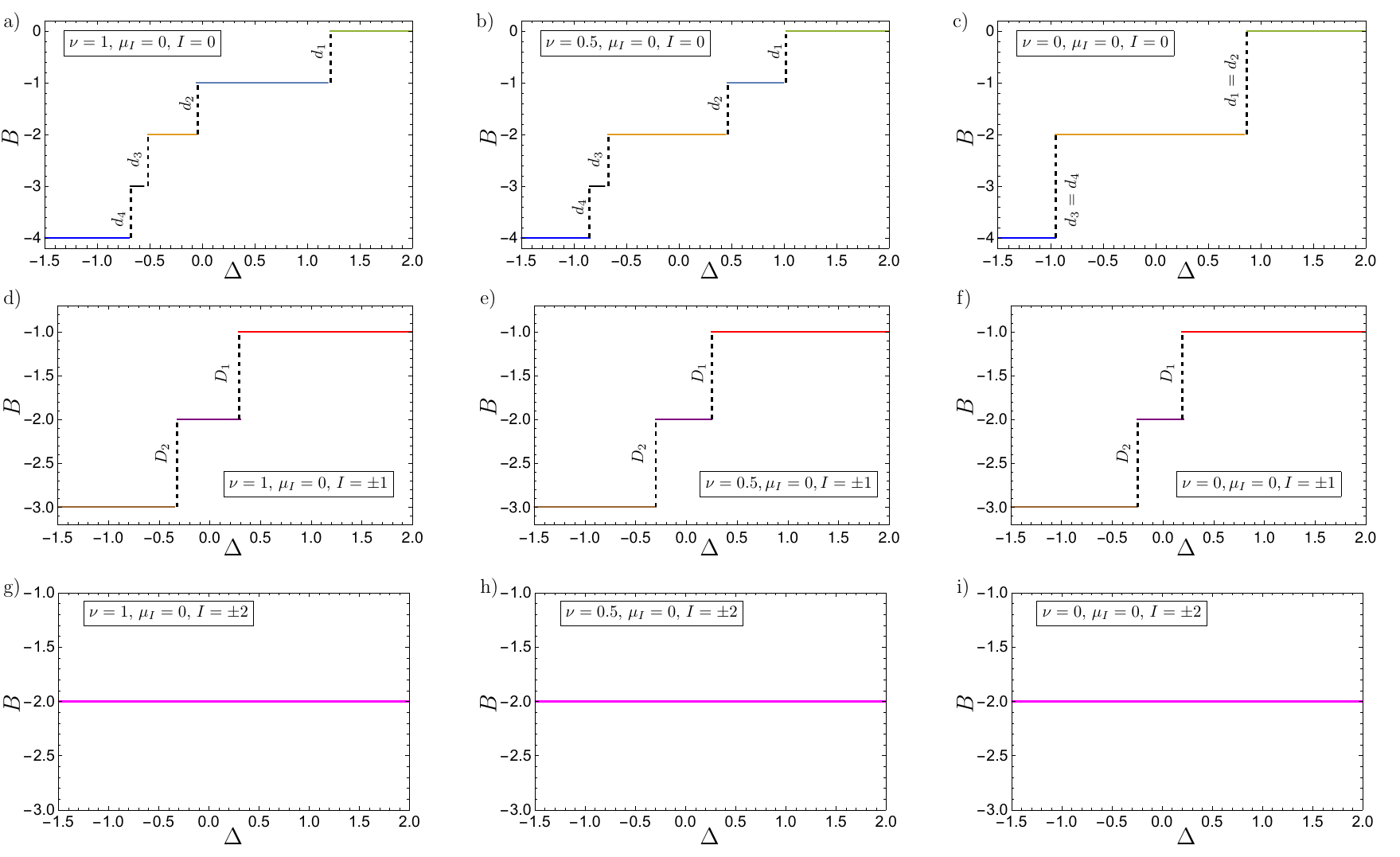}
\caption{ Baryon number $B$ of the lightest states with different $I$ as a function of $\Delta$ at ${\mu}_{_{I}}=0$ for various values of $\nu$. The dashed lines represent the locations of the level crossings.}\label{Fig6_sebsec_1}
\end{center}
\end{figure} 

\underline{${\mu}_{_{I}}=0$:} The energies of the lightest states $\mathcal{E}_0^{I}$ for different values of $I$ (with different possible  values of $B$)  are shown in Fig.~\ref{Fig2_sebsec_1} as functions of $\Delta$ and $\nu$.
As is evident, there is an intricate structure of level crossings and the corresponding quantum phase transitions.  
The number and locations of these level crossings depend on  $I$ and $\nu$.  

For the lightest state with $I=0$, there are four level crossings  (Fig.~\ref{Fig2_sebsec_1}a-c). The locations of these crossings $d_1, d_2, d_3$ and $d_4$ depend on the coupling $\nu$ as shown in Fig.~\ref{Fig_di_Di}a.  Near $\nu \simeq 1$, the four crossings are well separated. As the coupling strength increases, the separation between $d_1$ and $d_2$ decreases and at $\nu\simeq 0$,   $d_1$ coincides with $d_2$. The same holds for the separation between $d_3$ and $d_4$ as well.   These correspond to triple crossings in Fig.~\ref{Fig2_sebsec_1}c.

The lightest state  with $I=0$ always  has spin $J=0$, as shown in Fig.~\ref{Fig7_sebsec_1}(a-c). On the other hand,  its baryon number  jumps discontinuously at every level crossing (Fig.~\ref{Fig6_sebsec_1}(a-c)).

\begin{figure}[ht!]
\begin{center} 
\includegraphics[width=18cm]{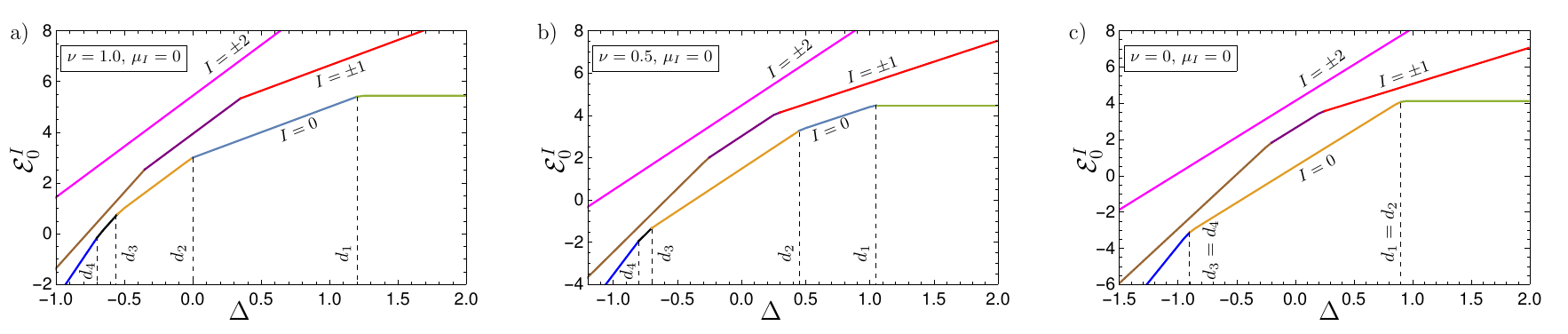}
\caption{Comparison of the lowest energy eigenvalues with different $I$ as a function of $\Delta$ at ${\mu}_{_{I}}=0$.  a) For $\nu=1$. b) For $\nu=0.5$. c) For $\nu=0$.  The different colored curves come from the lightest states in Fig.~\ref{Fig2_sebsec_1}. The spin and baryon number corresponding to each colored curve are shown in Fig.~\ref{Fig7_sebsec_1} and Fig.~\ref{Fig6_sebsec_1}. } \label{Fig1_sebsec_1}
\end{center}
\end{figure} 

 In the $I=\pm 1$ sector, there are only two level crossings in the lightest state (Fig.~\ref{Fig2_sebsec_1}d-f). The locations of these crossings $D_1$ and $D_2$ depend on $\nu$ (Fig.~\ref{Fig_di_Di}b) and they always remain well-separated. The spin of this lightest state does not change at the level crossings (Fig.~\ref{Fig7_sebsec_1}(d-f)) but its baryon number is discontinuous at $D_1$ and $D_2$ (Fig.~\ref{Fig6_sebsec_1}(d-f)). Interestingly, in this case, the lightest  state is  a spin-1 triplet.

Finally, for  $I=\pm 2$, the lightest state for all $\nu$ and $\Delta$ is devoid of level crossings and has  $J=0$ and $B=-2$  (Fig.~\ref{Fig7_sebsec_1}-\ref{Fig6_sebsec_1} (g-i)).

Fig.~\ref{Fig1_sebsec_1} shows that  there are no level crossings between the lightest states with different $I$.  The ground state for  ${\mu}_{_{I}} = 0$ is always a unique spin-0 state with $I=0$. However,  there are four level crossings at $d_1$, $d_2$, $d_3$ and $d_4$ in the $I=0$ sector and each of these crossings  corresponds to a QPT. There are thus five distinct phases, each  labelled by  its  $B$:
\begin{eqnarray}
 \text{ Phase I$_{B}$: \,\, when } d_k>\Delta> d_{k+1}, \quad\quad  B=-k, \quad\quad k=0,1,2,3,4, \quad\quad d_0=-d_5=\infty.  \label{cond_phaseI}
\end{eqnarray}
The first derivative of the ground state energy changes discontinuously across each $d_k$ and the QPTs are thus first order.

\begin{figure}[ht!]
\begin{center} 
\includegraphics[width=18cm]{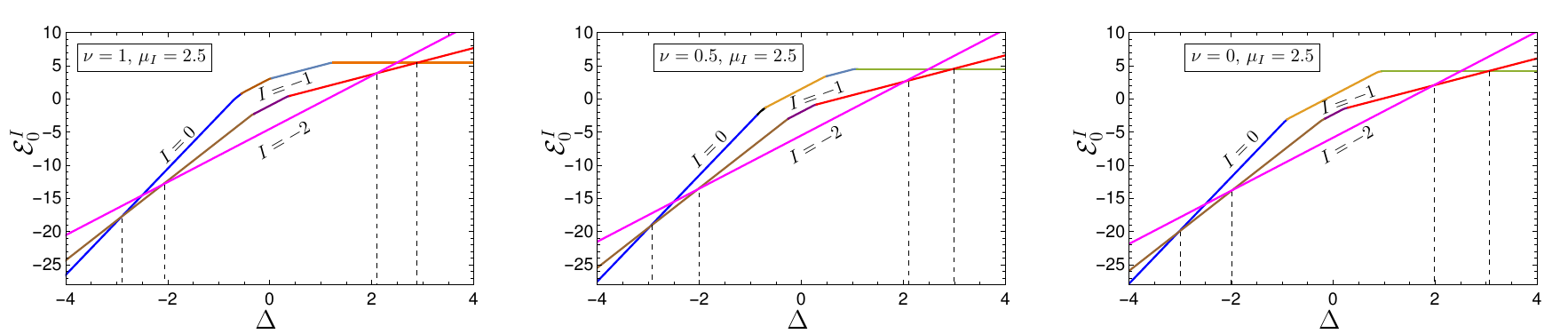}
\caption{Comparison of the lowest energy eigenvalues with different $I$ 
as a function of $\Delta$ at non-zero isospin chemical potential ${\mu}_{_{I}}=1.5$. The dashed vertical lines indicate the level crossings between states with different $I$.  
Left: $\nu=1.0$.  Center: $\nu=0.5$. Right: $\nu=0$. }
 \label{Fig3_sebsec_1}
\end{center}
\end{figure}

\underline{${\mu}_{_{I}} > 0$:} Now the energies  of the  states with different $I$ shift as
\begin{eqnarray}
\mathcal{E}_n^I({\mu}_{_{I}})=\mathcal{E}_n^I(0) +2 {\mu}_{_{I}} I. 
\end{eqnarray}
The energies in the $I=0$ sector remain unaltered, while for ${\mu}_{_{I}}>0$, the states with $I<0$  become lighter than their counterparts at ${\mu}_{_{I}}=0$.  As  ${\mu}_{_{I}}$ changes,  the lightest states with different $I$  can cross (which does not happen  when ${\mu}_{_{I}}=0$).

Indeed, several such level crossings do happen (see Fig.~\ref{Fig3_sebsec_1}) in the ground state when ${\mu}_{_{I}} \neq 0$. 
Thus the true ground  state energy $\mathcal{E}_{gs}({\mu}_{_{I}})$ is 
\begin{eqnarray}
    \mathcal{E}_{gs}(  \mu_{_I})= \text{min}\{\mathcal{E}_0^{(-2)}, \mathcal{E}_0^{(-1)},
    \mathcal{E}_0^{(0)},
    \mathcal{E}_0^{(1)}, \mathcal{E}_0^{(2)}\}.
\end{eqnarray}
The ground state can have $I=0$ (which we call phase I$_{B}$) or $I=- 1$ (which we call phase II$_{B}$) or $I=- 2$ (which we call phase III$_{B}$).  The level crossings in the ground state  correspond to QPTs between these phases. 

To locate these crossings, it is useful to define the energy gaps:  
 \begin{eqnarray}
    e_{_{01}} \equiv 2\Big[\mathcal{E}_0^{(\pm 1)} -\mathcal{E}_0^{(0)}\Big]_{{\mu}_{_{I}}=0}, \quad \quad 
    e_{_{02}} \equiv 4\Big[\mathcal{E}_0^{\pm 2} -\mathcal{E}_0^{0}\Big]_{{\mu}_{_{I}}=0}, \quad \quad
    e_{_{12}} \equiv 2 \Big[\mathcal{E}_0^{\pm 2} -\mathcal{E}_0^{\pm 1}\Big]_{{\mu}_{_{I}}=0}. 
\end{eqnarray}
 It is easy to see  that $e_{01}$,  $e_{02}$, and  $e_{12}$ are positive. 

If  ${\mu}_{_{I}}< e_{_{01}} $ and ${\mu}_{_{I}}< e_{_{02}} $, then the ground state has $I=0$ and its energy is $\mathcal{E}_{gs}({\mu}_{_{I}})=\mathcal{E}_0^0$. This case corresponds to phase I$_{B}$,  with conditions for different $B$ phases given in (\ref{cond_phaseI}).  

If ${\mu}_{_{I}}>  e_{_{02}}$ and ${\mu}_{_{I}}>  e_{_{12}}$, then the ground state has $I= -2 $. The ground state energy is $\mathcal{E}_{gs}({\mu}_{_{I}})=\mathcal{E}_0^{-2}$ and it corresponds to phase III$_{B}$. For $I= -2$, the only possible value of baryon number is $B=-2$ and hence there is no further splitting of this phase. 

\begin{figure}[ht!]
\begin{center} 
\includegraphics[width=18cm]{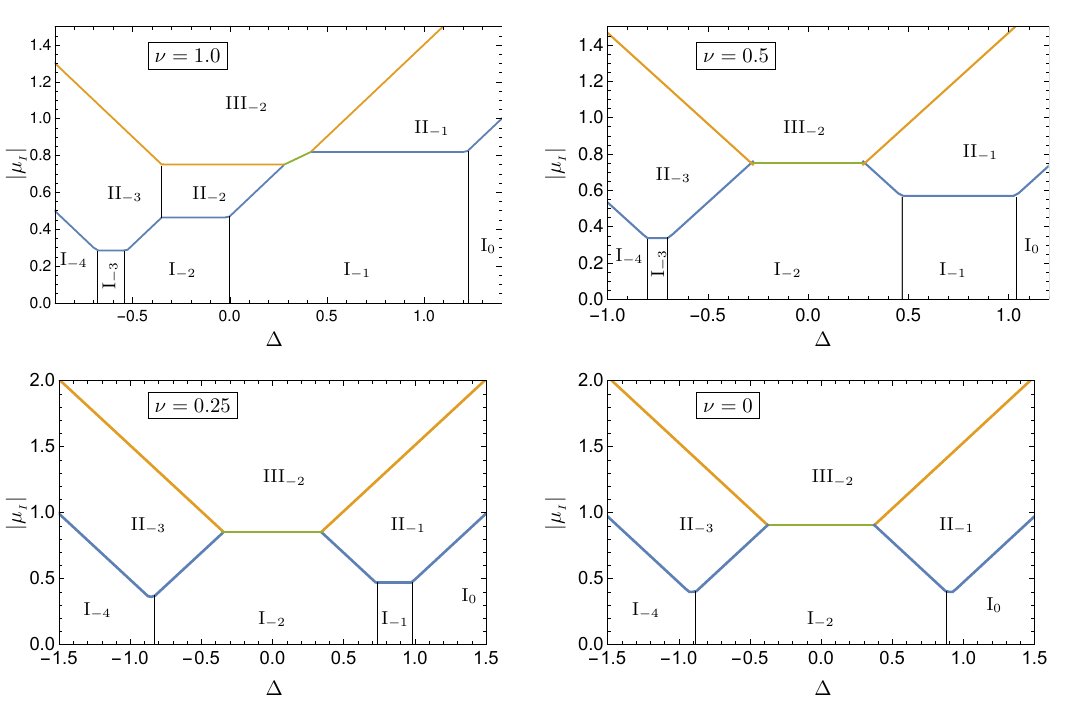}
\caption{ The phase diagram  in the $\Delta-|{\mu}_{_{I}}|$ plane for different $\nu$ when ${\mu}_{_{B}} \to \infty$ and ${c} \to \infty$.} \label{Fig4_sebsec_1}
\end{center}
\end{figure}

Finally, if $e_{01}<{\mu}_{_{I}}<  e_{12}$, then the  ground state has $I= -1$ with  $\mathcal{E}_{gs}({\mu}_{_{I}})=\mathcal{E}_0^{- 1}$. This corresponds to phase II$_{B}$, where $B=-3,-2,-1$. The condition for each of these phases with distinct baryon numbers is
\begin{eqnarray}
 \text{ Phase II$_{B}$: \,\, when } D_k>\Delta> D_{k+1}, \quad\quad  B=-(k+1), \quad\quad k=0,1,2, \quad\quad D_0=-D_3=\infty.  \label{cond_phaseII}
\end{eqnarray}
The phase diagram in the $\Delta-{\mu}_{_{I}}$ plane for different coupling regimes is shown in Fig.~\ref{Fig4_sebsec_1}.  The quantum numbers of the ground states of these phases are listed in Table~\ref{tab:phases_del1_muI}.

\begin{table}[h!]
\centering
\mbox{} \\ 
{\small 
\begin{tabular}{|c|c|c|c|c|c|}
\hline 
Phase & Conditions for   
& Isospin  & Baryon number & Spin & Degeneracy  \\ 
&  
  the phase  &$I$ &  $B$ & $J$ &\\ 
\hline \hline  & & &&& \\ 

I$_{B}$ & $|{\mu}_{_{I}}|<e_{_{01}},e_{_{02}}$ & 0 & $B=-4,-3,-2,-1,0$ & $0$ & $1$ \\  & && condition for different $B$ is (\ref{cond_phaseI}) &&  \\ 
 & &&&& \\
II$_{B}$ & $e_{_{12}}>|{\mu}_{_{I}}|>e_{_{01}}$ & $-sgn({\mu}_{_{I}})$ & $B=-3,-2,-1$   & $1$ & $3$ \\
 & && condition for different $B$ is (\ref{cond_phaseII}) &&  \\ 
 & &&&& \\
III$_{B}$ &$|{\mu}_{_{I}}|>e_{_{02}},e_{_{12}}$& $-2sgn({\mu}_{_{I}})$ & $B=-2$   & $0$ & $1$ \\  & &&&& \\

\hline
\end{tabular}
}\\  
\mbox{} \\ 
\caption{
The phases in the $\mu_{_B} \to \infty$ and $c \to \infty$ limit.}
\label{tab:phases_del1_muI}
\end{table}

Let us point out some salient features of the phases in Fig.~\ref{Fig4_sebsec_1}. The phases I$_{B}$, II$_{-3}$, II$_{-1}$ and III$_{-2}$ exist for all $\nu$. In the $\nu\to 0$ limit, the Hamiltonian is invariant under particle-hole exchange ($N_{d_1} \leftrightarrow  4-N_{d_1}$ and  $N_{d_2} \leftrightarrow  4-N_{d_2}$)  and $\Delta \leftrightarrow -\Delta$. In the phase diagram, this appears as  the reflection symmetry  about $\Delta=0$.  As $\nu \to 0$, the  phase regions of I$_{-1}$,  I$_{-3}$ and II$_{-2}$ shrink.  At $\nu=0$,  the phases I$_{-1}$ and I$_{-3}$  exist only on the phase boundaries, which correspond to the triple crossings in the ground state energy (see Fig.~\ref{Fig2_sebsec_1}c).  In contrast, the region of phase II$_{-2}$ which is trapezoidial in shape at $\nu \simeq 1$ (Fig.~\ref{Fig4_sebsec_1}a) shrinks as $\nu$ decreases and reduces to a point at $\nu\simeq 0.6$ (not shown in figure). For $\nu \lesssim 0.6$, this phase II$_{-2}$ does not exist.

\begin{figure}[ht!]
\begin{center} 
\includegraphics[width=7cm]{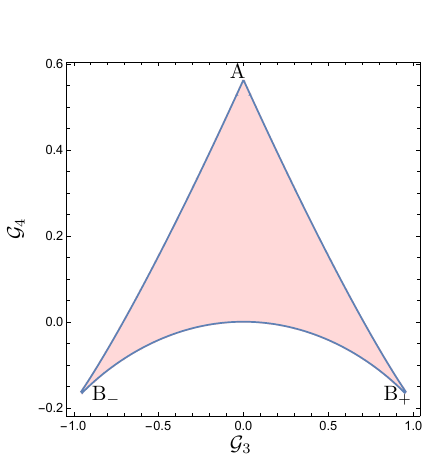}
\caption{The classically allowed region in  $(\mathcal{G}_3, \mathcal{G}_4)$ plane.} \label{Fig_arrow}
\end{center}
\end{figure}

The third and fourth order bosonic Binder cumulants  \cite{Acharyya:2024pqj} further reveal  important properties of these phases.    
These are defined as follows: 
\begin{eqnarray}
\mathcal{G}_3 \equiv \frac{\sqrt{3}}{2}  \epsilon_{ijk} \epsilon_{abc}  \frac{  \langle\Psi_{gs}| M_{ia} M_{jb} M_{kc}| \Psi_{gs} \rangle  } {\langle\Psi_{gs}| M_{ia} M_{ia} |\Psi_{gs} \rangle^{\frac{3}{2}} }, \quad\quad 
\mathcal{G}_4 \equiv \frac{9}{16}  \frac{ \langle \Psi_{gs}|(2M_{ib} M_{jc} M_{ic} M_{jb}-M_{ia} M_{ia} M_{jb} M_{jb})| \Psi_{gs} \rangle }{\langle \Psi_{gs}| M_{ia} M_{ia}| \Psi_{gs} \rangle^2}.  \label{g3_g4_defn} 
\end{eqnarray}
 The classical analogues of $\mathcal{G}_3$ and $\mathcal{G}_4$ are constrained to lie inside the ``arrowhead''  (shaded region of Fig.~\ref{Fig_arrow}). The interior (bulk) of the arrowhead corresponds to matrix configurations $M_{ia}$ with three distinct singular values, while the edges (AB$_\pm$ and $B_+B_-$) and corners (A and B$_\pm$) correspond to matrix configurations with degenerate (two or more identical) singular values  \cite{Pandey:2016hat, Acharyya:2024pqj}.   The tip A of the arrowhead with  $\mathcal{G}_3=0$ and $\mathcal{G}_4=\frac{9}{16}$ is particularly interesting:  it corresponds to matrix configurations with only one non-zero singular value.

\begin{figure}[ht!]
\begin{center} 
\includegraphics[width=16cm]{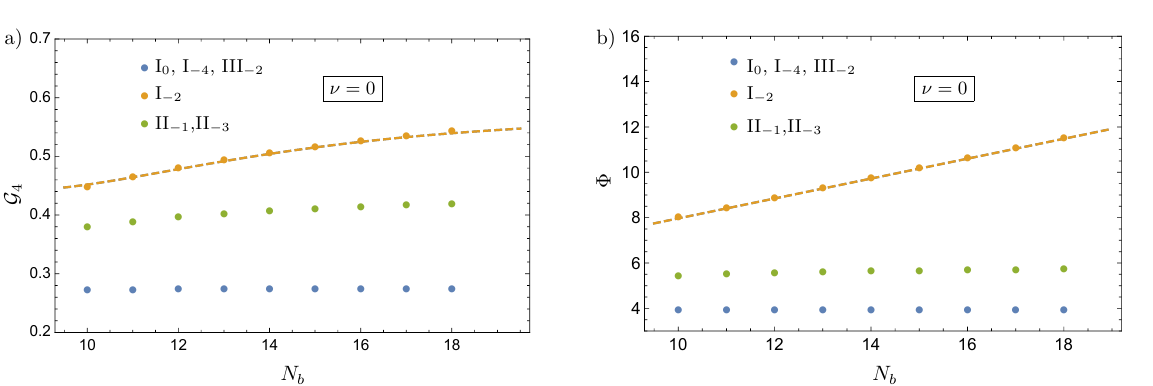}
\caption{ a) $\mathcal{G}_4$ and b) $\Phi$ for the phases at $\nu=0$ as functions of the bosonic cut-off $N_b$.  The dashed line in a) represents the fits in (\ref{fits})-(\ref{fit_parameters}). In b), the dashed line is the fit: $3.58 + 0.44 N_b$.  } \label{Fig_G4_Phi}
\end{center}
\end{figure}

The values of $\mathcal{G}_3$ and  $\mathcal{G}_4$ in these phases at different $\nu$ are given in Table~\ref{tab:phases_bosonic_obs}. 
For $\nu>0$, $(\mathcal{G}_3, \mathcal{G}_4)$ lies strictly in the interior of the arrowhead.  At $\nu=0$,  the phases in Fig.~\ref{Fig4_sebsec_1} have $\mathcal{G}_3 = 0$ because  the cubic term of the Hamiltonian vanishes, while $\mathcal{G}_4$ remains  non-zero (Fig.~\ref{Fig_G4_Phi}).  
\begin{table}[H]
\centering
\mbox{} \\ 
{\small 
\begin{tabular}{ |c||c|c|c||c|c|c||c|c|c||c|c|c| }
\hline 
& \multicolumn{3}{|c||}{$\nu=0$} 
& \multicolumn{3}{|c||}{$\nu=0.25$} 
& \multicolumn{3}{|c||}{$\nu=0.5$} 
& \multicolumn{3}{|c|}{$\nu=1.0$} \\ 
\hline
Phase 
& $\mathcal{G}_3$ & $\mathcal{G}_4$ & $\Phi$
& $\mathcal{G}_3$ & $\mathcal{G}_4$ & $\Phi$
& $\mathcal{G}_3$ & $\mathcal{G}_4$ & $\Phi$
& $\mathcal{G}_3$ & $\mathcal{G}_4$ &$\Phi$
\\
\hline \hline

I$_{-4}$
& 0 & 0.272 & 3.901
& 0.062 & 0.262 & 3.866
& 0.129 & 0.236 & 3.784
& 0.277 & 0.164 & 3.619
\\

I$_{-3}$
& --- & --- & ---
& --- & --- & ---
& 0.083 & 0.283 & 4.750
& 0.275 & 0.173 & 4.363
\\

I$_{-2}$
& 0 & $\frac{9}{16}$ & $\infty$
& 0.018 & 0.553 & 13.02
& 0.067 & 0.474 & 7.682
& 0.3 & 0.213 & 5.392
\\

I$_{-1}$
& --- & --- & ---
& 0.148 & 0.280 & 5.16
& 0.260 & 0.207 & 5.124
& 0.527 & 0.054 & 5.764
\\

I$_0$
& 0 & 0.272 & 3.901
& 0.062 & 0.262 & 3.866
& 0.129 & 0.236 & 3.784
& 0.277 & 0.164 & 3.619
\\

\hline

II$_{-3}$
& 0 & 0.435 & 5.794
& 0.05 & 0.394 & 5.465
& 0.120 & 0.318 & 4.891
& 0.322 & 0.164 & 4.390
\\

II$_{-2}$
& --- & --- & ---
& --- & --- & ---
& --- & --- & ---
& 0.429 & 0.104 & 5.057
\\

II$_{-1}$
& 0 & 0.435 & 5.794
& 0.05 & 0.394 & 5.465
& 0.120 & 0.318 & 4.891
& 0.322 & 0.164 & 4.390
\\

\hline

III$_{-2}$
& 0 & 0.272 & 3.901
& 0.062 & 0.262 & 3.866
& 0.129 & 0.236 & 3.784
& 0.277 & 0.164 & 3.619
\\

\hline
\end{tabular}
}\\  
\mbox{} \\ 

\caption{$\mathcal{G}_3$, $\mathcal{G}_4$ and $\Phi$ in the ground state of different phases  at the $N_b \to \infty$ limit. For absent phases, the corresponding entries are denoted by dashes.  }
\label{tab:phases_bosonic_obs}
\end{table}

As evident from the figure, $ \mathcal{G}_4$ for the phases I$_0$, I$_{-4}$, III$_{-2}$, II$_{-1}$,   and II$_{-3}$   converge for $N_b \gtrsim 16$: 
\begin{eqnarray}
 \mathcal{G}_4[{\text{I}_0}]= \mathcal{G}_4[{\text{I}_{-4}}]= \mathcal{G}_4[{\text{III}_{-2}}] \simeq 0.27, \quad\quad  \mathcal{G}_4[{\text{II}_{-1}}]= \mathcal{G}_4[{\text{II}_{-3}}]\simeq 0.42.  
\end{eqnarray}
The $(\mathcal{G}_3$, $\mathcal{G}_4)$ values for these phases lie in the interior of the arrowhead.

In contrast, for the phase I$_{-2}$ at $\nu=0$, $\mathcal{G}_4$ strongly depends on $N_b$, which may be fitted as 
\begin{eqnarray}
&& \mathcal{G}_4 [{\text{I}_{-2}}] \approx a \tanh(b \, N_b -d) + \frac{k}{N_b^\alpha},  \quad \text{where} \\
&& a \simeq 0.5615 \approx \frac{9}{16}, \quad b\simeq 0.1065, \quad d\simeq 0.338, \quad k \simeq 32.53, \quad \alpha \simeq \frac{5}{2}.  \label{fit_parameters}
 \label{fits}
\end{eqnarray}
From above, we get $\lim\limits_{N_b \to \infty} \mathcal{G}_4[{\text{I}_{-2}}]= a \approx \frac{9}{16}$ and 
  thus, the pair $(\mathcal{G}_3, \mathcal{G}_4)$ for the phase I$_{-2}$  at $\nu=0$ lies at the tip A. 
This is not unexpected as the Hamiltonian of  matrix-QCD$_{2,2}$ with large $\mu_B$ and $c$ but small $\Delta$ and $\mu_I$ effectively reduces to that of massless matrix-QCD$_{2,1}$ with small  chemical potentials. Indeed, the chiral limit of matrix-QCD$_{2,1}$  at $\nu=0$ exhibits a phase localized  at  the tip A \cite{Acharyya:2024pqj}. Here, however this phase exists irrespective of the quark mass.

Another interesting observable is $\Phi\equiv \langle \Psi_{gs}| M_{ia} M_{ia}| \Psi_{gs}\rangle$, whose values in different phases are given in Table~\ref{tab:phases_bosonic_obs}. It is easy to see that $\Phi$ jumps discontinuously at the phase boundaries. Further, we find that in phase I$_{-2}$, $\Phi$ diverges linearly with $N_b$: it can be fitted as $\Phi\simeq 3.58 +0.44 N_b$. A similar   (power law) divergence in $\Phi$    
was found in the weak coupling (large $\nu$) regime in adjoint QC$_2$D \cite{Acharyya:2026uhx} where it was argued  that the divergence signals the appearance of the non-regular representation of the Weyl algebra.  Here, the divergence in $\Phi$ is observed in the extreme strong coupling $(\nu=0)$. Whether the underlying physics is conceptually the same for these two situations is an interesting question and will be investigated elsewhere.

\subsubsection{Analogue of LOFF phases} 
 Among these phases, phase-II$_{B}$'s are particularly interesting, where  
 the ground states are spin-1 triplets with  $I=\mp 1$ when $\mu_{_I}$ is positive/negative.  Since the ground state carries non-zero spin, the rotational invariance is  broken. These phases  are reminiscent of the LOFF-like phases in two-color QCD \cite{Splittorff:2000mm, Fukushima:2007bj}. 

The usual LOFF phase has Cooper pairs with non-zero center-of-mass momentum, which defines a preferred direction in space and breaks the translational and rotational symmetry \cite{Larkin:1964wok,Fulde:1964zz}. In  QC$_2$D with non-zero chemical potentials, such phases have been predicted \cite{Splittorff:2000mm, Fukushima:2007bj}, where di-(anti-)quarks play the role of the Cooper pairs and carry momentum $\vec{q}$. There, the di-(anti-)quark condensate $\propto e^{-i \vec{q}\cdot \vec{x}}$ is the order parameter, which is spatially modulated (periodic) and hence corresponds to an inhomogeneous superconductor. As a consequence, both translational and rotational invariances are broken.

The matrix model is an approximation of the two-color QCD upon compactifying the spatial $\mathbb{R}^3$ to $S^3$ and projecting to the zero-momentum sector of the quark-glue dynamics. Now the question is whether the matrix model is  able to capture the LOFF phase and if yes, what are its signatures.  

Let us consider the zero-momentum limit in the usual description of the condensate in LOFF phase. This corresponds to the very long wavelength modulation of the condensate. As a result, the condensate remains nearly constant and thus is a homogeneous state. 
However, the direction $\widehat{q}$ remains well-defined even in this limit, and the limiting state is homogeneous but anisotropic, i.e. it breaks the rotational invariance.  Thus the signature of the LOFF phase in this limit is carried by the spontaneous  breaking of rotational symmetry.

This is what we see in the matrix model as well. The matrix model, being a $(0+1)-$dimensional approximation of Yang-Mills theory,  cannot capture any spatial modulation. However, as we show below, in phase-II$_{B}$, the spin of the ground state defines a preferred direction and thus breaks $SO(3)_{\rm rot}$. The fermionic part of this ground state has the properties of di-anti-quarks and its spin is dominantly polarized along that direction.

 To see the breaking of $SO(3)_{\rm rot}$ in these phases, 
let us introduce an external perturbation  
$\vec{\epsilon}\cdot\vec{J}$, with $\vec{\epsilon} \in \mathbb{R}^3$, 
and analyze the ground state in the  $|\vec{\epsilon}| \to 0$ limit. The perturbed Hamiltonian is
\begin{eqnarray}
    H_\epsilon= H+ \epsilon(\sin \theta \cos\varphi J_1+ \sin \theta \sin\varphi J_2 + \cos \theta J_3), \quad\quad \epsilon\geq 0, \quad 0\leq \theta\leq \pi, \quad\quad 0 \leq \varphi \leq 2\pi,
\end{eqnarray}
where $H$ is the unperturbed Hamiltonian  (\ref{Ham_1}-\ref{Ham_2}). Since $[H_\epsilon, J_i]\neq 0$ when $\epsilon\neq 0$, the perturbation breaks $SO(3)_{rot}$ explicitly. 

Let us denote the  ground state of these phases as $\{|\psi^+\rangle, |\psi^0\rangle, |\psi^-\rangle\}$ -- a spin-1 multiplet with $J_3=\{+1, 0,-1\}$. When $\epsilon=0$, these three states are degenerate with the ground state energy, say $E_{gs}$. The perturbed Hamiltonian in this triplet sector is simply
\begin{eqnarray}
    H_\epsilon= \left(
\begin{array}{ccc}
E_{gs}& 0 &0\\
0& E_{gs}& 0 \\
0 & 0 & E_{gs}
    \end{array}
\right)+ \epsilon
    \left(
\begin{array}{ccc}
  \cos \theta  & \frac{1} {\sqrt{2}}e^{-i \varphi } \sin \theta  & 0 \\
 \frac{1} {\sqrt{2}}e^{i \varphi } \sin \theta  & 0 & \frac{1} {\sqrt{2}}e^{-i \varphi } \sin \theta  \\
 0 & \frac{1} {\sqrt{2}}e^{i \varphi } \sin \theta  &  -\cos \theta  \\
\end{array}
\right). 
\end{eqnarray}
The ground state $|\Psi_{gs}\rangle$ of $H_\epsilon$ is unique: 
\begin{eqnarray}
    |\Psi_{gs}\rangle = e^{-i \varphi} \sin^2 \frac{\theta}{2} |\psi^+\rangle - \frac{1}{\sqrt{2}} \sin \theta |\psi^0\rangle+  e^{i \varphi} \cos^2 \frac{\theta}{2} |\psi^-\rangle.  
\end{eqnarray}
and its energy  is $E_{gs}- \epsilon$. 
Importantly the ground state energy depends on $\epsilon$, while the state depends only on $\theta$ and $\varphi$.  Thus in the limit of $\epsilon \to 0$, the ground state energy smoothly approaches $E_{gs}$ while the state continues to depend on $\theta$ and $\varphi$.  
Thus $SO(3)_{rot}$ is spontaneously broken by the ground state. 

In the $\epsilon\to 0$ limit, $J_3$ though still an observable is not a constant of motion. In fact, 
\begin{eqnarray}
    \langle J_3\rangle_{\text{phase-II}_{B}} \equiv \langle \Psi_{gs}|J_3|\Psi_{gs}\rangle_{\text{phase-II}_{B}} = - \cos \theta. 
\end{eqnarray}

This ground state contains a linear combination of composite fermion-glue states. The fermionic states therein are  either spin-0 or spin-1 states which transform  as singlets or triplets under the gauge rotations.  Thus, these fermionic states, which carry the baryon number and isospin, have the properties of a di-anti-quark (spin-$\frac{1}{2}\otimes \frac{1}{2}$ and color-$\frac{1}{2}\otimes \frac{1}{2}$).  This raises an interesting question: how much is the contribution of these  di-anti-quark-like states to $J_3$? 

We denote  these fermionic states as $|s,s_3\rangle_{_F}$ (with $s=0,1$ and $s_3=-s,0,s$). As the ground state has spin-1, the glue states in the composites have  spin 0,1 or 2 and we denote these glue states as $|\ell, \ell_3\rangle_{_G}$ (with $\ell=0,1,2$ and $\ell_3=-\ell, -\ell+1, \ldots \ell-1, \ell$). The spin triplet $\{|\psi^{+}\rangle,|\psi^{0}\rangle,|\psi^{-}\rangle\}$ can be expressed as
\begin{eqnarray}
|\psi^{J_3}\rangle = \sum_{(s,\ell )} c_{s\ell} \,\, C^{1, J_3}_{s, s_3; \ell, \ell_3} |s,s_3\rangle_{_F}\otimes |\ell, \ell_3\rangle_{_G},  \quad\quad J_3 =\pm 1, 0 
\end{eqnarray}
where $ C^{1, J_3}_{s, s_3; \ell, \ell_3}$ are the Clebsch-Gordan coefficients, $(s,\ell) =\{(1,0),(0,1),(1,1),(1,2)\}$, and the coefficients  satisfying  $|c_{01}|^2+ |c_{10}|^2+ |c_{11}|^2+|c_{12}|^2=1$ are determined by the ground state wavefunction. In above, the  color indices of the glue and fermionic states  are  suppressed for brevity. 

The total spin of the state gets contribution from both the fermions and the glue: $\vec{J}= \vec{S} + \vec{L}$. This gives the contribution of $S_3$ to $J_3$ of the ground state: 
\begin{eqnarray}
 \langle S_3\rangle_{\text{phase-II}_{B}} = -\Big(|c_{10}|^2+\frac{1}{2}|c_{11}|^2-\frac{1}{2}|c_{12}|^2\Big) \cos \theta \equiv - f_{_{B}} \cos \theta. 
\end{eqnarray}
Here, $f_{_{B}}$ denotes the fraction of $J_3$ carried by the di-anti-quark-like fermionic states. Not surprisingly, this fraction depends only on the contribution from $|1,s_3\rangle_{_F}$ and not on $|0,0\rangle_{_F}$. 

It is easy to see that $f_{_{B}}$ satisfies the bound  $-\frac{1}{2} \leq f_{_{B}}\leq 1$. When $f_{_B}=1$, the spin of the fermionic state is perfectly aligned with the spin of the ground state. When  $f_{_{B}}=-\frac{1}{2}$, the spin of the fermionic state is dominantly anti-aligned with the spin of the ground state. When $f_{_{B}}=0$,  there is no net alignment and the fermionic spin remains unpolarized.

\begin{figure}[ht!]
\begin{center} 
\includegraphics[width=15cm]{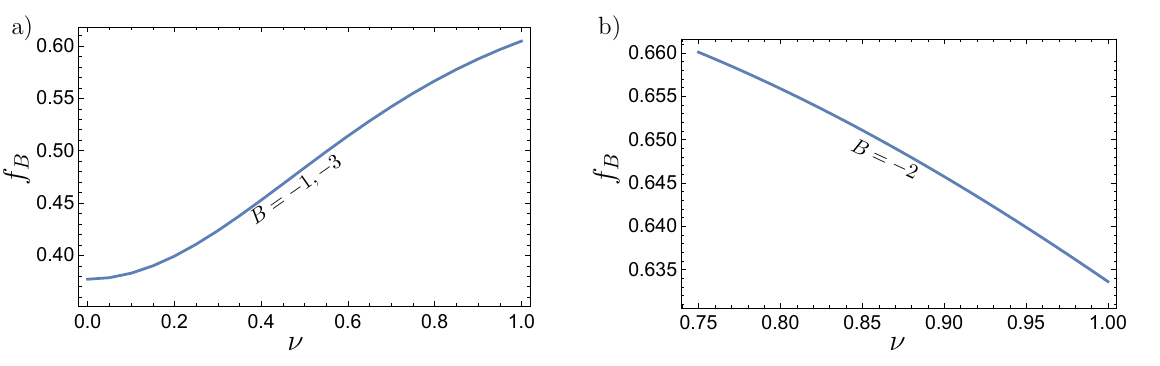}
\caption{The spin fractions $f_{B}$ vs $\nu$ in the phase-II$_{B}$. a) $B=-1$ and $B=-3$. b) $B=-2$. } \label{Fig8_sebsec_1}
\end{center}
\end{figure}

The actual estimates of $f_{_{B}}$ in phase-II$_{B}$ ground state  as a function $\nu$ (with $\Delta$ and $\mu_I$ appropriately chosen)  are shown in Fig.~\ref{Fig8_sebsec_1}. All the $f_{_{B}}$'s are positive  in these phases:  $f_{-2}(\nu) \gtrsim 0.63$ and $f_{-1}(\nu)=f_{-3}(\nu) \gtrsim 0.38$. Thus the spin of the fermionic state has a  net alignment along the total spin of the state. 

Thus we find the surprising result that  for any $\nu$, when $\mu_{_I}$ and $\Delta$ are suitable for phase II$_{B}$,  a spin-1 di-anti-quark state becomes lighter than its  mesonic counterpart.  There is an unexpected sharing of the total spin between fermionic and glue states, similar to the \cite{Acharyya:2024pqj}.

 \subsection{Large $|{\mu}_{_{B}}|$ and $|{\mu}_{_{I}}|$ }
In this subsection, we consider the regime in which ${\mu}_{_{B}}\to\infty$ and ${\mu}_{_{I}}\to\infty$, while $ c$ and the difference ${\mu}_{_{BI}} \equiv ({\mu}_{_{B}}-{\mu}_{_{I}})$ remain finite. The  same limit has been considered in \cite{Splittorff:2000mm} and we will reproduce several of their results.

This case is quite different from the ones discussed in Sec.~\ref{sec4_1}. Firstly, energies and the eigenstates depend on the quark mass $m$.  Secondly, for the low-lying energy eigenstates, the baryon number and the isospin are related (see Eqn.~(\ref{N_B_I_reln})): $B= -2 + \ell$ and $I=-2-\ell$ with $-2\leq \ell\leq 2$. The energy eigenstates $|\Psi_n^\ell \rangle$ and eigenvalues $\mathcal{E}_n^{(\ell)}$ are labelled by $\ell$: 
\begin{eqnarray}
    \mathcal{E}_n^{(\ell)}({\mu}_{_{BI}})=  \mathcal{E}_n^{(\ell)} (0) + 2 {\mu}_{_{BI}} \ell.
\end{eqnarray} 
For any given $ m$, $ c  $ and $\nu$, the energy eigenvalues obey $\mathcal{E}_n^{(\ell)}(0)=\mathcal{E}_n^{(-\ell)} (0)$.

\begin{figure}[ht!]
\begin{center} 
\includegraphics[width=15cm]{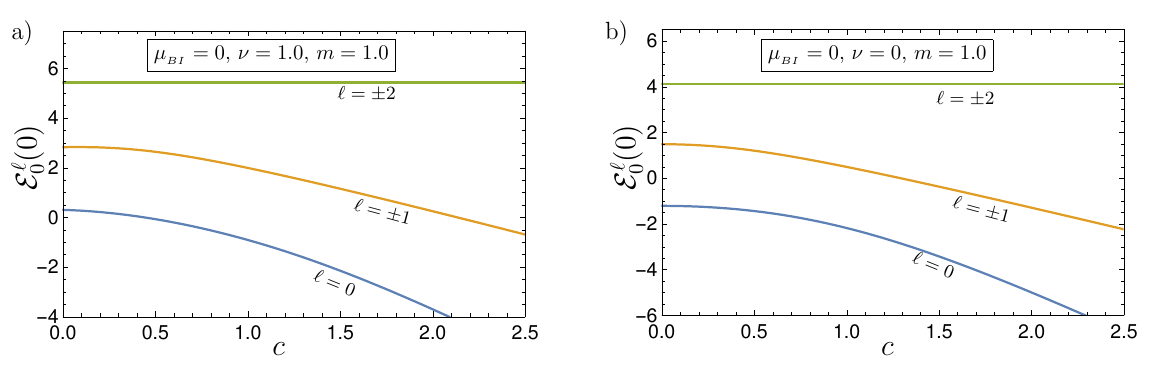}
\caption{ $\mathcal{E}_0^\ell(0)$ as a function of ${c}$ for fixed $m$. a) Intermediate coupling $\nu =1$ and b) extreme strong coupling $\nu=0$. In both cases, we have chosen $m=1.0$. } \label{Fig1_sebsec_2}
\end{center}
\end{figure}

The numerical estimates of $\mathcal{E}_0^\ell(0)$ for a given mass are shown in Fig.~\ref{Fig1_sebsec_2}. Here, we have chosen $m=1$ for the purpose of demonstration. 
Numerically we find that the lightest states with $\ell=0$ and $\ell=\pm 2$ is always  a spin-0 state (for all $\nu$, $ c$ and $ m$), while for $\ell=\pm 1$, it is a spin-1 multiplet. The true ground state is obtained by comparing the energies of these lightest states: $\mathcal{E}_{gs} = \text{min}\{\mathcal{E}_0^{(\ell)}: \, \ell=-2,-1,0,1,2\}$

\begin{figure}[ht!]
\begin{center} 
\includegraphics[width=15cm]{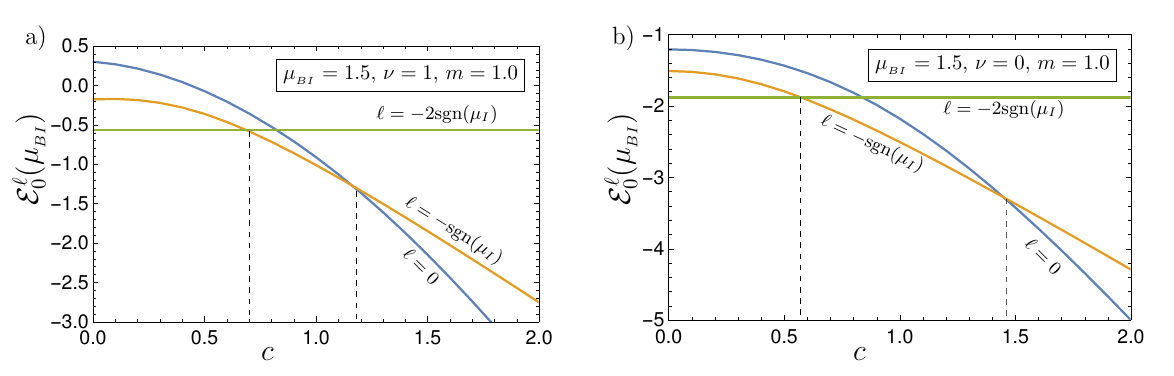}
\caption{$\mathcal{E}_0^\ell$ vs ${c}$ for fixed $m$ when ${\mu}_{_{BI}}$ is non-zero. a) Intermediate coupling $\nu=1$ and b) extreme strong  coupling $\nu=0$.  For both cases, we have chosen ${m}=1.0$ and ${\mu}_{_{BI}}=1.5$ for demonstration purposes.  } \label{Fig2_sebsec_2}
\end{center}
\end{figure}

When ${\mu}_{_{BI}}=0$, the ground state is $|\Psi_0^{0} \rangle$ -- the lightest state with $\ell=0$ for all $ c$, $\nu$ and $ m$. When ${\mu}_{_{BI}}$ is non-zero, there are possibilities of level crossing QPTs and the ground state can have $\ell\neq 0$ (see Fig.~\ref{Fig2_sebsec_2}). 
Across these level crossings, the baryon number and the isospin of the ground state jump discontinuously.  This gives different phases shown in Table~\ref{table_phases_2}. 
\begin{center}

{\small 
\begin{table}[h!]
\begin{tabular}{|c|c|c|c|c|c|}
\hline 
Phase & Conditions for the phase & Isospin  & Baryon number & Spin & Degeneracy  \\ 
&   &$I$ &  $B$ & $J$ & \\ 
\hline \hline  & & &&& \\ 

IV & $\left\{\begin{array}{cc}
    |2{\mu}_{_{BI}}|<\mathcal{E}_0^{(\pm 1)}(0) - \mathcal{E}_0^{(0)}(0) \\ \\
    |4{\mu}_{_{BI}}|<\mathcal{E}_0^{(\pm 2)}(0) - \mathcal{E}_0^{(0)}(0)
\end{array}\right.$
 & -2 & -2 & $0$ & $1$ \\  & &&  &&  \\ 
 & &&&& \\
V$_\pm$ & $\left\{\begin{array}{cc}
    |2{\mu}_{_{BI}}|>\mathcal{E}_0^{(\pm 1)}(0) - \mathcal{E}_0^{(0)}(0) \\ \\
    |2{\mu}_{_{BI}}|<\mathcal{E}_0^{(\pm 2)}(0) - \mathcal{E}_0^{(\pm1)}(0)
\end{array}\right.$
 & $-2-sgn({\mu}_{_{BI}})$ & $-2+sgn({\mu}_{_{BI}})$ & $1$ & $3$ \\  & &&  &&  \\ 
 & &&&& \\
VI$_\pm$ & $\left\{\begin{array}{cc}
    |2{\mu}_{_{BI}}|>\mathcal{E}_0^{(\pm 2)}(0) - \mathcal{E}_0^{(0)}(0) \\ \\
    |2{\mu}_{_{BI}}|<\mathcal{E}_0^{(\pm 2)}(0) - \mathcal{E}_0^{(\pm 1)}(0)
\end{array}\right.$
 & $-2-2sgn({\mu}_{_{BI}})$ & $-2+2sgn({\mu}_{_{BI}}) $ & $0$ & $1$ \\  & &&  &&  \\

\hline
\end{tabular}
\caption{The phases in the $\mu_{_B} \to \infty$ and $\mu_{_I} \to \infty$ limit. }\label{table_phases_2}
\end{table}
}
\end{center}

\begin{figure}[ht!]
\begin{center} 
\includegraphics[width=18cm]{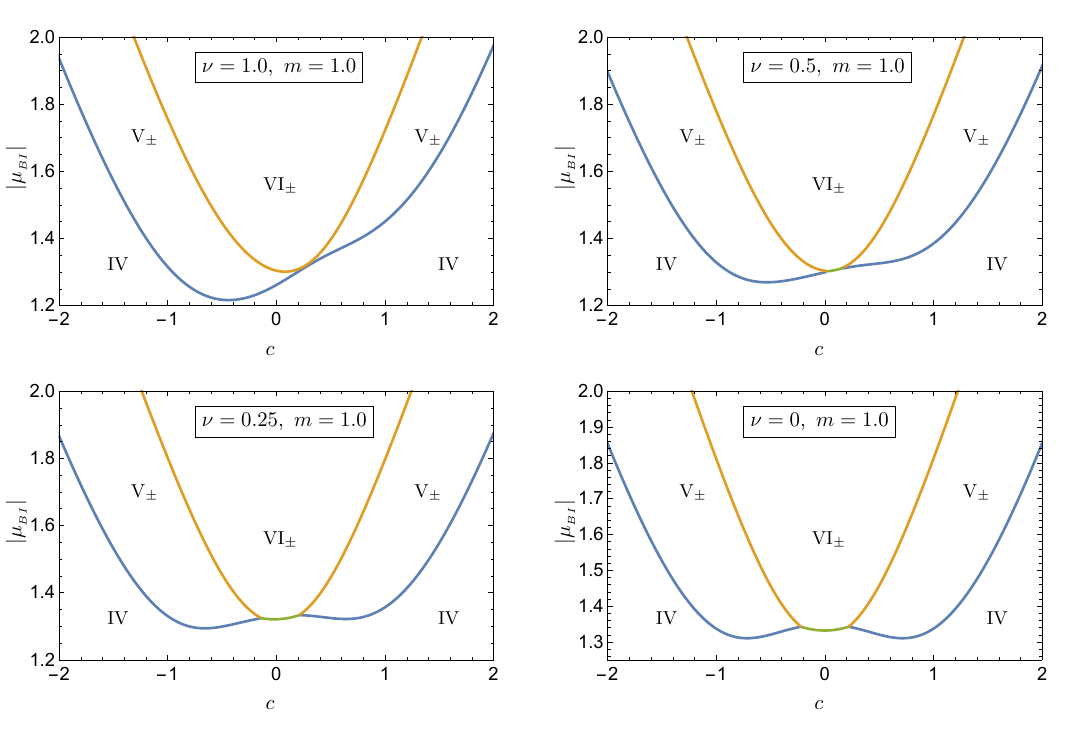}
\caption{The phase diagram in the $ c-|{\mu}_{_{BI}}|$ plane for different $\nu$ with fixed $ m=1$ when ${\mu}_{_{B}} \to \infty$ and ${\mu}_{_{I}} \to \infty$. } \label{Fig_phase_diag2}
\end{center}
\end{figure}

For a generic value of $ m$ (here we have chosen $ m=1$ as an example), the phase diagrams for various $\nu$ are shown in Fig.~\ref{Fig_phase_diag2}. {As $\nu$ decreases, the phase diagram progressively tends to become symmetric on reflection about the $c=0$ line at $\nu=0$. This is due to the exchange symmetry $b_{\alpha A f} \leftrightarrow d_{\alpha A f} $ and $c \leftrightarrow -c$ at $\nu=0$.  In the chiral limit (i.e. $m \to 0$), the phase diagrams remain qualitatively similar and all these phases persist. On the other hand, in the heavy quark limit  (i.e. $m \gg 1$), the phase regions of V$_\pm$ shrink for both intermediate and ultra-strong coupling.  Thus, in the heavy quark limit, the  phases IV and VI$_\pm$ become dominant.  }

 Again, these phases have different numerical values for $\mathcal{G}_3$, $\mathcal{G}_4$ and $\Phi$. Here, these observables depend not only on $\nu$ but also on $m$ and $c$,  and change discontinuously across the phase transitions.  For example when $m=1$ and $c=\frac{3}{2}$, their values are given in Table \ref{tab:phases_bosonic_obs_2}.  For this choice of $m$ and $c$, ($\mathcal{G}_3$, $\mathcal{G}_4$) lies in the interior of the arrowhead for all $\nu$ and the corresponding $\Phi$ is finite.   
\begin{table}[h!]
\centering
\mbox{} \\
{\small
\begin{tabular}{ |c||c|c|c||c|c|c||c|c|c||c|c|c| }
\hline
& \multicolumn{3}{|c||}{$\nu=0$}
& \multicolumn{3}{|c||}{$\nu=0.25$}
& \multicolumn{3}{|c||}{$\nu=0.5$}
& \multicolumn{3}{|c|}{$\nu=1.0$} \\
\hline

Phase
& $\mathcal{G}_3$ & $\mathcal{G}_4$ & $\Phi$
& $\mathcal{G}_3$ & $\mathcal{G}_4$ & $\Phi$
& $\mathcal{G}_3$ & $\mathcal{G}_4$ & $\Phi$
& $\mathcal{G}_3$ & $\mathcal{G}_4$ & $\Phi$
\\
\hline \hline

IV
& $0.01$ & $0.30$ & $4.28$
& $0.08$ & $0.28$ & $4.24$
& $0.15$ & $0.24$ & $4.14$
& $0.35$ & $0.14$ & $4.14$
\\

V$_\pm$
& $0.00$ & $0.39$ & $5.24$
& $0.05$ & $0.36$ & $5.02$
& $0.12$ & $0.30$ & $4.62$
& $0.32$ & $0.16$ & $4.23$
\\

VI$_\pm$
& $0.00$ & $0.27$ & $3.90$
& $0.06$ & $0.26$ & $3.87$
& $0.13$ & $0.24$ & $3.78$
& $0.28$ & $0.16$ & $3.62$
\\

\hline
\end{tabular}
}\\  
\mbox{} \\

\caption{$\mathcal{G}_3$, $\mathcal{G}_4$, and $\Phi$ for phases IV, V$_\pm$, and VI$_\pm$ at different values of $\nu$ with $m=1$ and $c=\frac{3}{2}$ at $N_b \to \infty$ limit.} \label{tab:phases_bosonic_obs_2}
\end{table}

\begin{figure}[ht!]
\begin{center} 
\includegraphics[width=18cm]{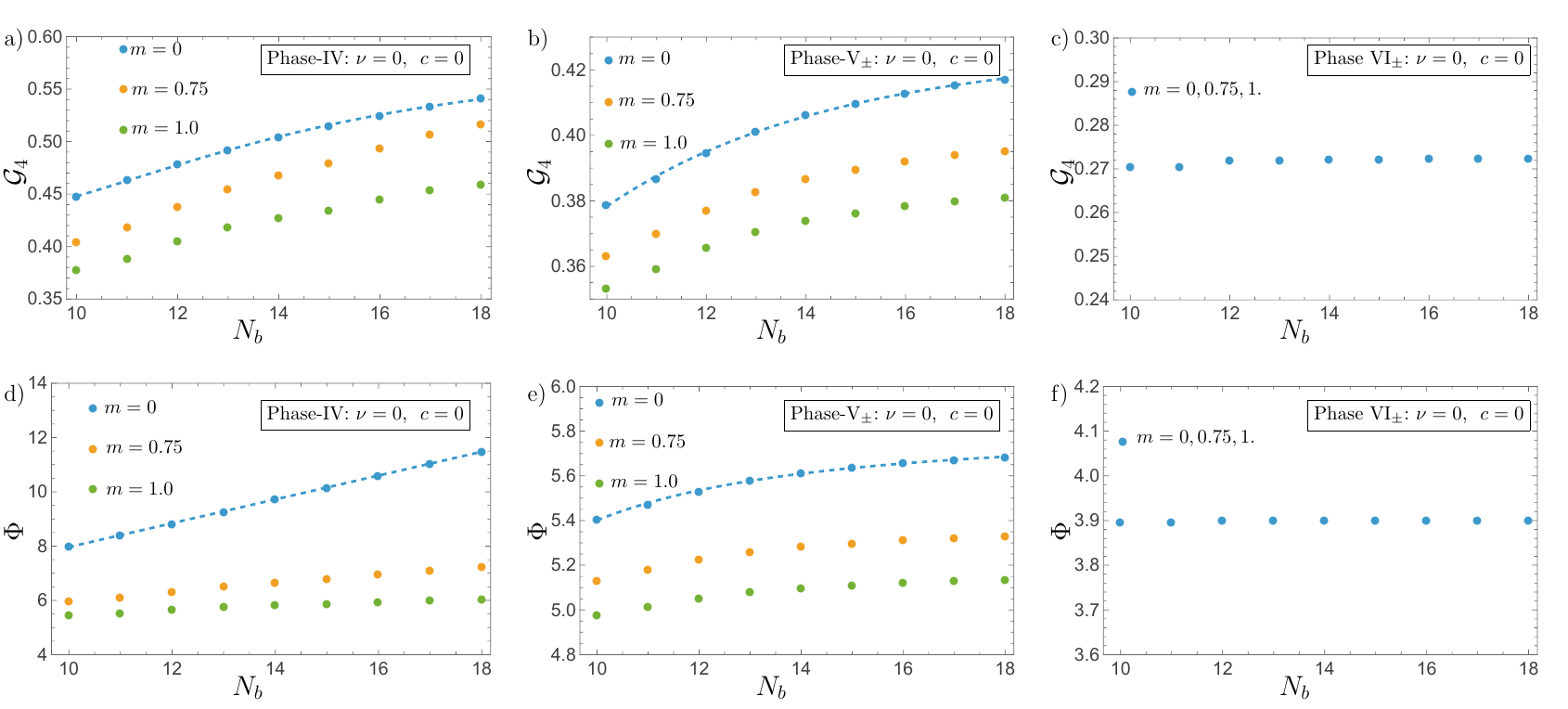} 
\caption{ a-c) The $\mathcal{G}_4$  and d-f) $\Phi$ in different phases at $\nu=0$ and $c=0$ for various $N_b$.  The dots correspond to the numerical data. The dashed lines in  a) and b) correspond to fits  (\ref{fit_AB})-(\ref{fit_parameter_AB}).  The dashed line in  d) is the linear fit $3.58 + 0.44 N_b$. In e), the dashed line represents the fit in (\ref{fit_phi_V}). } \label{Fig_G4_phi_phaseABC}
\end{center}
\end{figure}

As a different example, we have plotted the $\mathcal{G}_4$ and $\Phi$ at $c=0$ and $\nu=0$ for various $m$ in Fig.~\ref{Fig_G4_phi_phaseABC}. As evident from the figure,  $\mathcal{G}_4[{\text{VI}_\pm}] \simeq 0.27 $ (and $\Phi [{\text{VI}_\pm}]\simeq 3.9 $) is independent of $m$ and remains in the interior of the arrowhead (and finite).

On the other hand, in phase IV and V$_\pm$,  both $\mathcal{G}_4$  and $\Phi$ depend on $m$. In the massless limit,  $\mathcal{G}_4$ in both these phases behaves as 
\begin{eqnarray}
\lim_{m \to 0} \mathcal{G}_4 \approx a \tanh(b \, N_b -d) + \frac{k}{N_b^\alpha} \label{fit_AB}
\end{eqnarray} 
where 
\begin{eqnarray}
&&  \text{Phase IV:} \,\,\,\, \quad a\simeq 9/16, \quad b\simeq 0.103, \quad d\simeq 0.294, \quad \,\,\,  k \simeq 15.19, \quad  \alpha \simeq 2.2, \nonumber  \\
&& \text{Phase V$_\pm$:} \quad \,\,\, a\simeq  0.43, \quad b\simeq 0.119, \quad  d\simeq -0.571, \,\,  \,\,\, k \simeq -2.17,  \,\,\, \,\alpha \simeq 1.9. \label{fit_parameter_AB}
\end{eqnarray}
Thus, in the $N_b \to \infty$ limit,  for phase V$_\pm$ with massless quarks  at $\nu=0$,   $(\mathcal{G}_3, \mathcal{G}_4)  \approx (0, 0.43)$  lies in the interior of the arrowhead.  Further, $\Phi$ in this phase remains finite:  as a function of $N_b$, it can be fitted as
\begin{eqnarray}
\Phi \simeq 5.78 \tanh \,(0.05 N_b +1.91) -\frac{226.7}{N_b^{2.9}}. \label{fit_phi_V}
\end{eqnarray}

In contrast for  phase IV with $m=0$,  $(\mathcal{G}_3, \mathcal{G}_4)  \approx (0,\frac{9}{16}) $ and  again $\Phi \simeq 3.58 + 0.44 N_b$.  As $m$ becomes non-zero and/or $c$ becomes significantly large, $(\mathcal{G}_3, \mathcal{G}_4)$ moves away from the  tip A,  and the divergence in $\Phi$ disappears.

\textit{LOFF phases:} In the ${\mu}_{_{B}}\to \infty$ and  $ {\mu}_{_{I}}\to \infty$ regime (different from the previous subsection) as well, there are LOFF-like phases. In particular in  phase V$_\pm$, the ground state is a spin-1 multiplet and $SO(3)_{rot}$ is again spontaneously broken. 
In this case too, the ground state depends on the direction along which the external perturbation $\vec{\epsilon}\cdot\vec{J}$ is tuned to zero and $ \langle J_3\rangle_{\text{V}_{\pm}}=-\cos \theta $. 

{
The baryon numbers of the ground states in these phases are $B=-1$ (or $B=-3$) for positive $\mu_{_{BI}}$  (or negative  $\mu_{_{BI}}$). The fermionic part of the ground state behaves like a diquark (when $B=-1$) or di-anti-quark (when $B=-3$), which has spin 0 or 1. It is straightforward to see that here also, the fermionic spin is $ \langle S_3\rangle_{\text{V}_{\pm}}= -  F(\nu, {m})\, \cos \theta$, where $F$ denotes the spin-fraction.   However, unlike in the previous subsection, here the spin-fraction is identical for all possible values of $B$ and depends on the quark mass $m$.

As before, the value of spin-fraction is theoretically bounded: $-\frac{1}{2}\leq F \leq 1$. The  actual values of $F$  as a functions of $\nu$ and ${m}$ are shown in Fig.~\ref{fig_F_vs_m_nu}. Again, the positive values of $F$ indicate a dominant alignment of the fermionic spin with the total spin of the ground state.

In the large $m$ limit, we expect the contribution of quark-glue interaction in the ground state energy is very small. This in turn implies that a spin-1 di-(anti-)quark $\otimes$ spin-0 glueball is an approximate eigenstate of $H$. If sufficiently large $\mu_{_B}$ and $\mu_{_I}$ are added, such a state can become the ground state. It is satisfying to see that the numerics confirms this intuition: 
$F \to 1$ as $m \to \infty$. 

\begin{figure}[ht!]
    \centering
    \includegraphics[width=18cm]{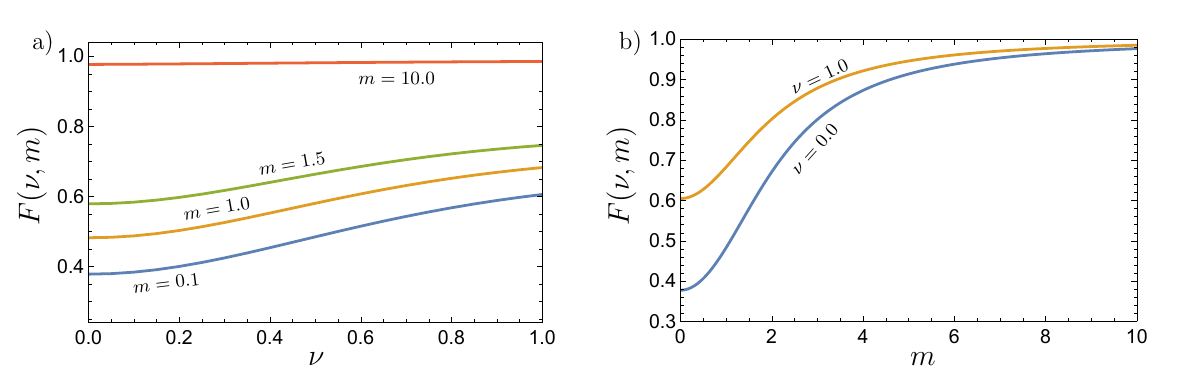}
    \caption{The ratio $F(\nu, {m})$ as a function of a) $\nu$ for fixed ${m}$ and b)  ${m}$ for fixed $\nu$.}
    \label{fig_F_vs_m_nu}
\end{figure}

 }

\section{Discussion}

It is not surprising that with quarks of multiple flavors,  the phase structure of the ground state is quite complicated. However, since real world QCD does have many flavors, it is obligatory to perform such investigations.  An important aspect of our work here is the reproduction of the LOFF phases argued by \cite{Splittorff:2000mm}.  Additionally, we have shown the existence of an exceptionally rich phase structure,  with multiple LOFF-like phases appearing in both intermediate and strong coupling regimes. Our framework  allows a systematic exploration of these phases at large chemical potentials, revealing  multiple phases in which the ground state has non-zero spin and spontaneously breaks  $SO(3)_{rot}$. Each such phase is characterized by an unique pair of quantum numbers:  baryon number $B$ and isospin $I$. 
For these spin-1 ground states, we find that the contribution of the quark spin (to the total spin) depends on the coupling strength and/or quark mass. In several situations, the quark spin, coming from a di-quark state, is the dominant component. 

In the present work, we have stayed away from the weak coupling regime.  Evidence from other investigation \cite{Acharyya:2026uhx} strongly suggests that in this regime, there can be a localization-delocalization transition, and the  approximations used in the current work can become unreliable. An investigation of this model without such  approximations is in progress and the results will be presented elsewhere.

Two phases warrant particular attention: the phase I$_{-2}$ and  the massless phase IV in the strong coupling where $\Phi$ diverges  and $(\mathcal{G}_3, \mathcal{G}_4)$ lies on the top tip of the arrowhead. A strikingly similar result has been discussed in \cite{Acharyya:2026uhx}, where in the weak coupling regime $(\mathcal{G}_3, \mathcal{G}_4)$  lies in the bottom right corner $B_+$ of the arrowhead. In that case, the wavefunction is translationally invariant  which leads to a divergent $\Phi$. These suggest that the wavefunction belongs to a non-regular representation of the Weyl algebra. The physical interpretations of these non-regular representations in the context of Yang-Mills theory need  to be better understood.

The diversity of the phases in this model suggests that the techniques applied here can perhaps be  used to understand the bewildering variety of phases of three-color QCD at low temperatures.

\appendix 
\section*{Appendices} 
\numberwithin{figure}{section}

\section{Rotational and Gauge symmetries of the Hamiltonian}\label{app_sym}

{The glue Hilbert space $\mathcal{H}_G=\displaystyle{L^2(M_3(\mathbb{R}),\prod_{ia} dM_{ia})}$ is infinite dimensional and the glue states may be organized in representations of spin and color.  These glue states always have integer spin and transform in odd-dimensional representations of the color $SU(2)$.  On the other hand, the quark Hilbert space $\mathcal{H}_F$ is  $2^{16}$-dimensional. The states with even number  fermions have integer spin and transform in odd-dimensional representations of color $SU(2)$. In contrast, the states with odd number of fermions have half-integer spin and transform in even-dimensional representations of color $SU(2)$.  The total Hilbert space is $\mathcal{H}_G \otimes \mathcal{H}_F$, where  the states may also be labelled by spin and color.

In the total Hilbert space $\mathcal{H} = \mathcal{H}_G \otimes \mathcal{H}_F$, 
the spatial rotations on the glue and the quark are generated by
\begin{eqnarray}
&&L_i \equiv - \epsilon_{ijk}\,P_{ja} M_{ka},
\qquad
S_i \equiv \frac{1}{2}\left(
b^\dagger_{\alpha A f}\sigma^i_{\alpha\beta} b_{\beta A f}
+ d_{\alpha A f}\sigma^i_{\alpha\beta} d^\dagger_{\beta A f}
\right),\\
&&[L_i, L_j] = i \epsilon_{ijk} L_k, 
\qquad 
[S_i, S_j] = i \epsilon_{ijk} S_k, \quad\quad i,j,k=1,2,3
\end{eqnarray}
where $\alpha,\beta=1,2$ denote spin indices, $A=1,2$ is the color index, 
and $f=1,2$ labels the flavor.

The gauge rotations on the glue and the quark sectors are generated by the Gauss-law generators
\begin{eqnarray}
&&G^a_{glue} \equiv -i \epsilon_{abc} M_{ib} P_{jc},\quad\quad G^a_{quark}\equiv 
b^\dagger_{\alpha A f} T^a_{AB} b_{\alpha B f}
+ d_{\alpha A f} T^a_{AB} d^\dagger_{\alpha B f},\\
&&[G^a_{glue},G^b_{glue}] = i\epsilon_{abc} G^c_{glue},\quad [G^a_{quark},G^b_{quark}] = i\epsilon_{abc} G^c_{quark},\quad a,b,c=1,2,3
\end{eqnarray}
The Hamiltonian (\ref{Ham_1}) does not commute with the above mentioned operators. It only commutes with the total spin $J_i=L_i+S_i$ and $G^a=G^a_{glue}+G^a_{quark}$:
\begin{eqnarray}
[J_i,J_j]=i\epsilon_{ijk}J_k,\quad [G^a,G^b]=i\epsilon_{abc}G^c,\quad [H,J_i]=0=[H,G^a]
\end{eqnarray}

Thus, the total angular momentum $J_i$ generates the spatial rotation group 
$SO(3)_{\mathrm{rot}}$, while the operators $G^a$ generate the $SU(2)$ gauge symmetry.

The Gauss-law constraint requires that all physical observables commute with $G^a$, and the physical Hilbert space $\mathcal{H}_{\mathrm{phys}}$ is the color-singlet subspace of $\mathcal{H}$. Therefore, any physical state $|\Psi\rangle \in \mathcal{H}_{\mathrm{phys}}$ satisfies
$G^a |\Psi\rangle = 0$. 

It is easy to see that  colorless states must have even number of quarks.  With two flavors, the number of quarks in any state cannot be greater than $16$. As a result, the color-singlet energy eigenstates have integer spin ($J=0,1,2 \ldots$), with $-4 \leq I\leq 4$ and  $-4 \leq B \leq 4$.

\end{document}